\documentclass[11pt]{article}
\pdfoutput=1
\usepackage[centertags]{amsmath}
\usepackage[square, comma, sort&compress,numbers]{natbib}
\usepackage{array,multirow}
\numberwithin{equation}{section}
\usepackage{amssymb,amsfonts}
\usepackage{graphicx}
\usepackage{color}
\usepackage{mathtools,bm}
\usepackage{accents}
\usepackage{amsmath}
\usepackage{cancel}

\usepackage{epsfig}
\usepackage{bbold}
\newcommand{\pref}{\gamma}
\usepackage{wrapfig}
\usepackage{float}
\usepackage{soul}
\usepackage{tkz-euclide}
\usepackage{braket}
\usepackage{tikz,pgf}
\usetikzlibrary{shapes}
\usetikzlibrary{calc}
\usetikzlibrary{decorations.pathmorphing}
\usetikzlibrary{decorations.pathreplacing,shapes.misc}
\usetikzlibrary{positioning}
\usetikzlibrary{arrows}
\usetikzlibrary{decorations.markings}
\usetikzlibrary{shadings}

\usetikzlibrary{intersections}

\newcommand{\be}{\begin{equation}}
\newcommand{\ee}{\end{equation}}
\newcommand{\ba}{\begin{array}}
\newcommand{\ea}{\end{array}}

\newcommand{\dps}{\displaystyle}
\newcommand{\half}{\frac{1}{2}}

\newcommand{\F}{\;_2F_1}

\newtheorem{prop}{Proposition}
\newtheorem{lemma}[prop]{Lemma}

\newtheorem{corollary}{Corollary}

\newcommand{\bref}[1]{\textbf{\ref{#1}}}

\newcommand{\im}{\mathop{\mathrm{Im}}}
\newcommand{\re}{\mathop{\mathrm{Re}}}

\newcommand{\CC}{\mathbb{C}}

\newcommand{\RR}{\mathbb{R}}

\newcommand{\ZZ}{\mathbb{Z}}

\newcommand{\cC}{\mathcal{C}}

\newcommand{\cO}{\mathcal{O}}

\newcommand{\cV}{\mathcal{V}}

\numberwithin{equation}{section} \makeatletter
\@addtoreset{equation}{section}

\newcommand{\ads}{AdS$_2\;$}

\newcommand{\h}{h}

\newcommand{\dth}{h}

\newcommand{\coef}{\kappa}

\newcommand{\sltwo}{sl(2,\mathbb{R})}

\newcommand{\bx}{{\bf x}}

\usepackage{jheppub}
\makeatletter
\def\@fpheader{\vspace{-.1cm}}
\makeatother

\title{\centering{Wilson  network decomposition of AdS Feynman diagrams in two dimensions}}

\author[a,b]{Konstantin\ Alkalaev}  
\author[a]{and Vladimir\ Khiteev}

\affiliation[a]{I.E. Tamm Department of Theoretical Physics, \\P.N. Lebedev Physical
Institute, 119991 Moscow, Russia}
\affiliation[b]{Institute for Theoretical and Mathematical Physics,\\
Lomonosov Moscow State University,
119991 Moscow, Russia}

\emailAdd{alkalaev@lpi.ru}
\emailAdd{khiteev@lpi.ru}

\abstract{We show that Feynman diagrams in AdS$_2$ space can be decomposed into  infinite series  of matrix elements of Wilson line network operators.  The case of the 3-point scalar Feynman diagram with endpoints in the bulk is studied in detail. The resulting decomposition is similar  to the conformal block decomposition of Witten diagrams, i.e. it comprises a single-trace term and infinite sums of double-trace terms. We derive a number of AdS propagator  identities which relate the standard bulk-to-bulk propagators with the modified bulk-to-bulk propagators of two different types responsible for extracting single-trace and double-trace terms.}

\begin{document}

\maketitle
\flushbottom

\section{Introduction}

\begin{figure}
\centering
\includegraphics[scale=0.63]{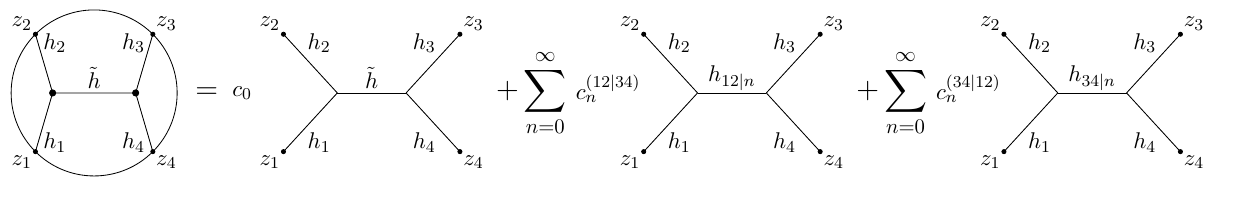}
\caption{The conformal block decomposition of the 4-point exchange Witten diagram. On the left-hand side: there are four  scalars of masses $m_i^2 = h_i(h_i-1)$, $i=1,...,4$, and one exchange scalar   of mass  ${\tilde m}^2 = \tilde h (\tilde h-1)$; the boundary points $z_i$; the two central dots denote the AdS integrations; the straight lines are  the standard bulk-to-bulk and bulk-to-boundary scalar propagators. On the right-hand side:  each term in this sum is the 4-point  scalar conformal block with specified intermediate conformal dimensions $\tilde h$ and $\dth_{ij|n} = h_i+\h_j+2n$, the $c$-coefficients are given in section \bref{sec:wit}. }
\label{fig:Witten_conf}
\end{figure}

Feynman diagrams in AdS space are built in the same way as in Minkowski space. All the basis principles of representing particles and their interactions with lines and vertices remain intact, with the only significant exception being that each particle can propagate in three distinct regimes: bulk-to-bulk, bulk-to-boundary, boundary-to-boundary. Of course, this multiplication of propagation types is directly related to the existence of the conformal boundary of AdS space endowed with the causal structure. This phenomenon underlies the AdS/CFT correspondence, where AdS Feynman diagrams with boundary points known as Witten diagrams calculate CFT correlation functions \cite{Maldacena:1997re,Gubser:1998bc,Witten:1998qj}. It follows that the large-$N$ CFT correlators   admit a holographic expansion in terms of Witten diagrams, a representation that is directly related to the conformal block decomposition. For example, in the case of the Euclidean AdS$_2$/CFT$_1$ correspondence, the 4-point exchange Witten diagram decomposed into conformal blocks is shown in fig. \bref{fig:Witten_conf}.

In this paper, we focus on  AdS Feynman diagrams which  have all their endpoints in the bulk, i.e. those which are composed from the bulk-to-bulk propagators only, and argue that these diagrams  have  a similar decomposition in terms of Wilson line networks.\footnote{Wilson  lines and their  networks  have many interesting properties, e.g. in the context of the AdS/CFT correspondence \cite{Ammon:2013hba,deBoer:2013vca,deBoer:2014sna,Hegde:2015dqh,Bhatta:2016hpz,Besken:2016ooo,Besken:2017fsj,Hikida:2017ehf,Anand:2017dav,Hikida:2018eih,Hikida:2018dxe,Besken:2018zro,Bhatta:2018gjb,DHoker:2019clx,Castro:2018srf,Kraus:2018zrn,Blommaert:2018oro,Hung:2018mcn,Castro:2020smu,Alkalaev:2020yvq,Belavin:2022bib,Belavin:2023orw,Castro:2023bvo,Alkalaev:2023axo}: near the conformal boundary they reproduce conformal blocks, in the  bulk they are   probes of the background geometry. } More precisely, one considers Feynman diagrams of scalar fields in two-dimensional Euclidean AdS space and shows that they can be decomposed into infinite sums of particular  matrix elements of Wilson line  networks stretched in the same space. Wilson  line  networks averaged between particular states  are purely group-theoretical objects obtained by multiplying Wilson line operators in AdS and 3-valent intertwiners (i.e. 3-j Wigner symbols) and in this respect  they are pretty similar to conformal blocks in CFT.\footnote{Recall that conformal blocks can be defined by averaging a particular combination of 3-valent intertwiners (this defines a channel) between primary states \cite{Moore:1988qv}.  }     

\begin{figure}
\centering
\includegraphics[scale=0.63]{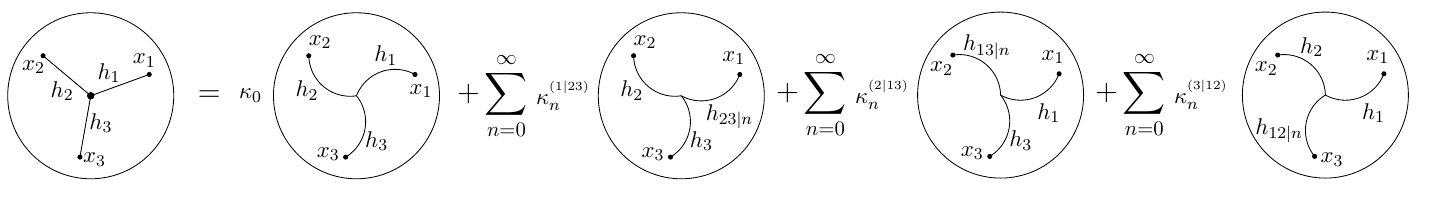}
\caption{The Wilson line network decomposition of the 3-point \ads Feynman  diagram. On the left-hand side: there are three  scalars of masses $m^2_i = h_i(h_i-1)$ located in the bulk  points $x_i$, $i=1,2,3$; the central dot denotes the AdS integration; the straight lines are the  standard bulk-to-bulk scalar propagators. On the right-hand side:  3-valent graphs denote matrix elements of Wilson line networks which are Wilson lines (wavy lines) of three different weights connected by $\sltwo$ intertwiners; weights $\dth_{ij|n} = h_i+\h_j+2n$, the $\coef$-coefficients are given in section \bref{sec:wilson_basis}.}
\label{fig:GGG_as_V}
\end{figure}

Our basic example is the 3-point AdS Feynman diagram  shown in fig. \bref{fig:GGG_as_V}. This decomposition closely resembles the  conformal block expansion  of Witten diagrams. While  the 3-point AdS Feynman diagram recovers the 3-point conformal correlation function on the boundary, its bulk structure reveals  a series  of terms remarkably similar to those  of the 4-point Witten diagram. Specifically, it comprises a ``single-trace'' term and infinite sums of ``double-trace'' terms. Furthermore, it is noteworthy that the 2-point AdS Feynman diagram -- representing the bulk-to-bulk scalar propagator -- is equivalent to the 2-point Wilson line network (i.e., an ordinary Wilson line in AdS space)  \cite{Castro:2018srf,Alkalaev:2024cje}.

There are two key points of our construction. The first one is that the HKLL  holographic reconstruction \cite{Hamilton:2005ju, Kabat:2011rz, Kabat:2015swa,Kabat:2020nvj} maps one-to-one  $n$-point conformal blocks on the boundary onto  Wilson network matrix elements  with $n$ endpoints in the bulk \cite{Alkalaev:2024cje}.\footnote{The 4-point CFT correlation function with one of fields being HKLL reconstructed was considered in \cite{Kabat:2016zzr}.} This is the general statement, although here we restrict ourselves to the 3-point case. We show that the 3-point  Wilson network matrix element  can be represented as an AdS integral involving a contour integration of  the threefold product of appropriately modified bulk-to-bulk propagators. This integral representation stems directly from the HKLL reconstruction formula.  The second one is the use of particular AdS propagator identities which allow one to transform the original AdS Feynman diagram into an infinite sum of AdS integrals over the modified bulk-to-bulk propagators, each of which is the Wilson network matrix element of particular conformal weights. The derivation and analysis of these propagator identities is central in this paper.

In section \bref{sec:HKLL} we briefly review the HKLL holographic reconstruction  which is needed for relating conformal blocks and Wilson line networks, here we introduce our conventions and notation. We also discuss  the conformal block decomposition shown in fig. \bref{fig:Witten_conf} as well as the AdS integral representation of conformal blocks in terms of geodesic Witten diagrams  obtained by applying specific  propagator identities that involve  the standard bulk-to-bulk(boundary) propagators. In section \bref{sec:new} we uplift the mentioned  propagator identities to AdS space. Here, we list a number of AdS propagator identities which relate    the standard bulk-to-bulk and modified bulk-to-bulk propagators, we provide motivation for their introduction and discuss their properties.  In section \bref{sec:prop} we introduce Wilson line networks and formulate several lemmas, which formalize the AdS propagator identities discussed in the previous section. Here, we show how  using these identities one can decompose  the 3-point AdS Feynman diagram into Wilson network matrix elements as shown in fig. \bref{fig:GGG_as_V}. Finally, we conclude with some discussions and future directions in section \bref{sec:conclusion}.   Appendix  \bref{app:Witten_analytic}  considers the analytic continuation of the 3-point Witten diagrams to the complex plane in the case of general and special dimensions. Appendix \bref{app:lemmas} contains the proofs of  lemmas formulated in section \bref{sec:prop}.      

\section{HKLL reconstruction and  Witten diagrams}
\label{sec:HKLL}

\subsection{Reconstruction formula}
\label{sec:reconst}

The HKLL reconstruction formula for the AdS$_{d+1}$ fields in terms of the boundary CFT$_d$ operators was derived in  a series of papers \cite{Hamilton:2005ju, Kabat:2011rz, Kabat:2015swa,Kabat:2020nvj}. Below we give the Euclidean version of the reconstruction formula in the AdS$_2$/CFT$_1$ case, as the AdS integrals  converge without the $\varepsilon$-prescription required in the Lorentzian case. 
Namely, assuming the $1/N$ expansion, the bulk reconstruction formula for the Euclidean AdS$_2$ scalar fields $\phi(\bx)$ of the mass $m^2=h(h-1)$ with $\bx=(u,z)$ reads  
\be 
\label{field_reconst}
\ba{l}
\dps
\phi(\bx) = \int_{\cC}dw\;\mathbb{K}_h(\bx,w)\, \cO_h(w) +
\vspace{2mm}
\\
\dps
\hspace{25mm} + \frac1N \sum_{k,l=0}^{\infty}\,\sum_{n=0}^{\infty}\frac{a(\h;\h_k,\h_l;n)}{\beta_{\h\h_k\h_l}}\int_{\cC}dw\;\mathbb{K}_{\dth_{kl|n}}(\bx,w)\,\cO_{\dth_{kl|n}}(w)+O\Big(\frac{1}{N^2}\Big)\,.
\ea
\ee 
 
\begin{enumerate}

\item The Poincare patch of the Euclidean AdS$_2$ in the local coordinates $\bx$ has the metric   
\be 
\label{metric}
ds^2 = \frac{dz^2+du^2}{u^2}\,,
\qquad
u\in\mathbb{R}_{\geq 0}\,,\; z \in \mathbb{R}\,.
\ee
The conformal boundary of the global \ads is disconnected. Here, we choose  one connected component,  which we  denote as  $\partial$AdS$_2$. It is placed at $u=0$. Also, the metric determinant $g(\bx) = 1/u^4$. 

\item $\cC$ is a line contour between the points $z\pm iu$ on the complexified conformal boundary, which is the standard complex plane, ($\partial$AdS$_2$)$^\mathbb{C} \cong \mathbb{C}$.  The integration contour $\cC$ is the Wick rotated integration contour $\cC_{\text{Lor}}$ of the Lorentzian HKLL integral, where $\cC_{\text{Lor}} = [z-u, z+u]$ is the line segment on $\RR$. It has a simple geometric interpretation: $\cC_{\text{Lor}}$ consists of boundary points which are spacelike separated from the bulk point $(u,z)$ in  the Lorentzian \ads. It follows that in the Euclidean HKLL integral formula \eqref{field_reconst} we deal with functions analytically continued on the complex plane, $w\in \mathbb{C}$.   

\item $\mathbb{K}_h(\bx,w)$ is the smearing function, which is the bulk-to-boundary propagator of dual conformal weight $1-h$:
\be
\label{smear}
\mathbb{K}_h(\bx,w) = \frac{-2i}{4^{\h}}\frac{\Gamma(2\h)}{\Gamma(\h)\Gamma(\h)}\left(\frac{u}{u^2 + (z-w)^2}\right)^{1-\h}.
\ee 
Note that our definition of the smearing function slightly differs from that  in \cite{Hamilton:2005ju}. The original HKLL formula integrates over the entire conformal boundary, and its smearing function uses Heaviside step functions to restrict the integration domain to the line contour $\cC_{\text{Lor}}$. Since the integration contour $\cC$ in the Euclidean version is complex, it is more convenient to resolve Heaviside step functions so that the smearing function is given by \eqref{smear}.

\item $\cO_h(w)$ is a primary  field  of  conformal weight  $h$. $\cO_{\dth_{kl|n}}(w)$ are double-trace fields built of two primary fields  $\cO_{h_k}$ and $\cO_{h_l}$ as
\be 
\cO_{\dth_{kl|n}}(w) = \sum_{i=0}^{2n}c_i(h_k,h_l,n)\,\partial^i \cO_{h_k}(w)\,\partial^{2n-i} \cO_{h_l}(w)\,,
\ee 
where  coefficients $c_i(h_k,h_l,n)$ can be chosen so that the double-trace field  is primary of  conformal weight  $\dth_{kl|n} = h_k+\h_l+2n$. The sum over $k$ and $l$ on the right-hand side of \eqref{field_reconst} goes over all primary fields  in a boundary CFT.  
    
\item The coefficients $\beta_{\h\h_k\h_l}$ are the 3-point structure constants in CFT. The coefficients $a(\h;\h_k,\h_l;n)$ (in the $d=1$ case) are given by 
\be 
\label{a_coef}
a(\h;\h_k,\h_l;n)=
\frac{2\pi^{\half}}{\Gamma(\h)}\,\frac{(-)^n(\h_k)_n(\h_l)_n}{n!(\h_k+\h_l-\half+n)_n} \,\frac{\Gamma(\h+\half)}{\h(\h-1)-(\dth_{kl|n})(\dth_{kl|n}-1)}\;,
\ee 
where $(a)_n = \Gamma(a+n)/\Gamma(a)$ is the Pochhammer symbol. They are fixed by   analyticity of the bulk-to-boundary correlation functions $\big\langle\phi(\bx)\cO_{\h_k}(w_2)\cO_{\h_l}(w_3)\big\rangle$ \cite{Hamilton:2005ju, Kabat:2011rz, Kabat:2015swa,Kabat:2020nvj}.

\end{enumerate}

\noindent The reconstruction formula \eqref{field_reconst} implies that the  corresponding  reconstruction relations  between  correlation functions of the bulk fields $\phi(\bx)$ and correlation functions of the primary fields $\cO_h(z)$ are represented in terms of multiple integrals of the boundary correlation functions which can be expanded in conformal blocks.  In general, such relations can be quite complicated. Near the conformal boundary, they are, however, simple: the extrapolate dictionary relation equates  the bulk correlation functions at $u\to0$ with the boundary conformal correlation functions \cite{Banks:1998dd}.  This can be directly seen from the reconstruction formula \eqref{field_reconst}: at  $u\to 0$ the integration contour $\cC$    shrinks to the boundary point $z$ that removes the integration; then, one shows that the second term is sub-leading, whereas  the leading term is proportional to the primary field, thereby producing the asymptotic value $\phi(\bx) \simeq u^h\cO_h(z)$.

\subsection{Conformal block decomposition}
\label{sec:wit}

\begin{figure}
\centering
\includegraphics[scale=0.7]{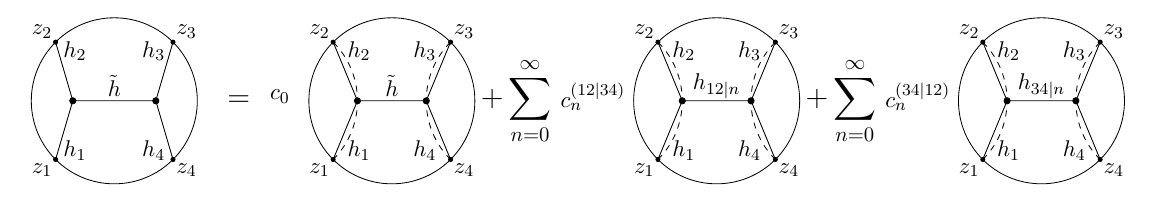}
\caption{The exchange Witten diagram \eqref{4pt_expansion} of four scalars   of mass $m_i^2 = h_i(h_i-1)$ and one exchange scalar   of mass  ${\tilde m}^2 = \tilde h (\tilde h-1)$. The conformal boundary $\partial$\ads is shown by the circle, the two bulk points are $\bx$ and $\bx'$. Each term in this decomposition is a geodesic Witten diagram obtained by restricting the integration domain of the exchange Witten diagram to geodesics (dashed  lines) connecting boundary points $z_1, z_2$ and $z_3, z_4$. Each geodesic Witten diagram computes the conformal block with particular intermediate conformal weight  \cite{Hijano:2015zsa}.}
\label{fig:witten_ex}
\end{figure}

Any bulk correlation function can be  decomposed into AdS Feynman  diagrams comprising multiple integrals over AdS space, where the integrands are products of bulk-to-bulk propagators. For a given diagram with all points on the boundary the resulting expression can be explicitly obtained as a sum of conformal blocks, where intermediate  channels correspond to single-trace and double-trace exchange fields inherited from the reconstruction formula \eqref{field_reconst} (note that at $O(1/N^2)$ only tree-level diagrams contribute). The example of the $4$-point exchange Witten diagram in AdS$_2$  space    reads (see fig. \bref{fig:witten_ex}) \cite{Hijano:2015zsa}\footnote{The $4$-point exchange Witten diagram was  calculated  in \cite{DHoker:1998ecp,DHoker:1999mqo}. The corresponding conformal block decomposition was found in \cite{Liu:1998th,Hoffmann:2000mx,Dolan:2001tt}. The 4-point contact Witten diagram has the similar representation \cite{Hijano:2015zsa}.}   
\be 
\label{4pt_expansion}
\ba{l}
\dps
\iint_{\text{AdS}_2}d^2\bx\,d^2\bx'\sqrt{g(\bx)g(\bx')} \; K_{\h_1}(\bx,z_1) K_{\h_2}(\bx,z_2)G_{\tilde{\h}}(\bx,\bx')K_{\h_3}(\bx',z_3)K_{\h_4}(\bx',z_4)=  
\vspace{3mm}
\\
\dps
\hspace{5mm}= c_0\, F_{h\tilde{h}}(z_1,...,z_4) + \sum_{n=0}^{\infty}c^{^{(12|34)}}_n\, F_{hh_{12|n}}(z_1,...,z_4)+ \sum_{n=0}^{\infty}c^{^{(34|12)}}_n\, F_{hh_{34|n}}(z_1,...,z_4)\,,
\ea
\ee 
where $G_h(\bx,\bx')$ is the bulk-to-bulk propagator of a scalar field of mass $m^2 = h(h-1)$ \cite{Fronsdal:1974ew}
\be 
\label{bulk-to-bulk}
G_h(\bx,\bx') = \left(\frac{\xi(\bx,\bx')}{2}\right)^h{}_2F_1\left[\frac{\h}{2},\frac{\h}{2}+\half; \h+\half\Big|\,\xi(\bx,\bx')^2\right],\quad \xi(\bx,\bx')= \frac{2u u'}{u^2+u'^2+(z-z')^2}\,;
\ee
the bulk-to-boundary propagator $K_h(\bx,z')$ is obtained by taking  one of the bulk points to the conformal boundary  
\be 
\label{bulk-to-boundary}
K_h(\bx,z') = \left(\frac{u}{u^2 + (z-z')^2}\right)^\h,
\ee 
cf. \eqref{smear}; the functions $F_{hh'}(z_1,...,z_4)$ are the  $4$-point scalar conformal blocks  with  external dimensions $h_1, ..., h_4$ and intermediate dimensions $h'=\tilde h$ or $h'=\dth_{ij|n} = h_i+h_j+2n$, and 
\be 
\label{coef_4pt}
\ba{l}
\dps
c_0 =  \beta_{\h_1\h_2\tilde{h}}\beta_{\tilde{h}\h_3\h_4}\sum_{m,n=0}^{\infty}\frac{a(\tilde{h};\h_1,\h_2;m)}{\beta_{\dth_{12|m}\h_1\h_2}}\,\frac{a(\tilde{h};\h_3,\h_4;n)}{\beta_{\dth_{34|n}\h_3\h_4}}\,,
\vspace{3mm}
\\
\dps
c^{^{(ij|kl)}}_n =  a(\tilde{h};\h_i,\h_j;n)\, \beta_{\dth_{ij|n}\h_k\h_l}\,\sum_{m=0}^{\infty}\frac{a(\dth_{ij|n};\h_k,\h_l;m)}{\beta_{\dth_{kl|m}\h_k\h_l}}\;,
\ea
\ee 
where the $a$-coefficients  are given by \eqref{a_coef}. It is worth noting that the $a$-coefficients arise in two different calculations: the HKLL reconstruction and the tree-level Witten diagram decomposition into conformal blocks, the relation between which has not been directly studied in the literature.   

There are three crucial steps in deriving the decomposition \eqref{4pt_expansion} \cite{Hijano:2015zsa,Hijano:2015qja}. Firstly, the infinite sum over single-trace and double-trace terms on the right-hand side is directly obtained  by using particular propagator identities applied to the bulk-to-bulk and bulk-to-boundary propagators in the integrand on the left-hand side. In particular, the identity producing the infinite sum on the right-hand side of \eqref{4pt_expansion} is given by\footnote{Here we integrated the identity from \cite{Hijano:2015zsa} with a test function.} 
\be
\label{geodesic_propid1}
\ba{c}
\dps
\int_{\text{AdS}_2}d^2\bx\sqrt{g(\bx)}\;K_{h_1}(\bx,z_1)K_{h_2}(\bx,z_2) f(\bx)= 
\sum_{n=0}^{\infty}\frac{1}{\beta_{\dth_{12|n}h_1h_2}} \frac{(-)^n(h_1)_n(h_2)_n}{n!(h_1+h_2+n-\half)_n}
\vspace{3mm}
\\
\dps
\times \int_{\gamma_{12}}d\lambda \int_{\text{AdS}_2}d^2\bx \sqrt{g(\bx)}\; K_{h_1}(\bx(\lambda),z_1)K_{h_2}(\bx(\lambda),z_2) G_{\dth_{12|n}}(\bx(\lambda), \bx)f(\bx)\,,
\ea
\ee
where $f(\bx)$ is a test function and $\gamma_{12}$ is a geodesic $\bx = \bx(\lambda)$ connecting two boundary points $z_1$ and $z_2$. Secondly, each term in the resulting infinite sum is given by the so-called geodesic Witten diagram, see fig. \bref{fig:witten_ex}. These diagrams have the same topology as the original diagram but now the integrals are taken over the geodesics instead of AdS$_2$:
\be
\iint_{\text{AdS}_2}d^2\bx\,d^2\bx' \sqrt{g(\bx)g(\bx')} \;\;\rightarrow\;\; \int_{\gamma_{12}} d \lambda \int_{\gamma_{34}} d \lambda' \,.
\ee
Thirdly, the geodesic Witten diagram computes the conformal block. In other words, the conformal block being a part of the conformal correlation function has its own AdS integral representation.\footnote{Geodesic Witten diagrams in AdS$_d$ were extensively studied from many perspectives, see e.g. \cite{Hijano:2015rla, Nishida:2016vds, Chen:2017yia, Castro:2017hpx, Dyer:2017zef, Sleight:2017fpc, Tamaoka:2017jce, Nishida:2018opl, Parikh:2019ygo, Jepsen:2019svc}. The analogous construction exists in AdS$_3$, where the large-$c$ Virasoro conformal blocks are represented by lengths of geodesic networks, see e.g.  \cite{Fitzpatrick:2014vua,Hartman:2013mia,Asplund:2014coa,deBoer:2014sna,Hijano:2015rla,Alkalaev:2015wia,Alkalaev:2015lca,Hijano:2015qja,Alkalaev:2016rjl,Banerjee:2016qca,Alkalaev:2016ptm,Alkalaev:2018nik}.} 

This method was applied to expand the $5$-point  and $6$-point contact and exchange Witten diagrams in different channels \cite{Parikh:2019ygo,Jepsen:2019svc}. While the $6$-point conformal block in the snowflake channel has a simple AdS representation, which directly generalizes the 4-point case, a similar  expression for the 5-point conformal block in the comb channel is more intricate.\footnote{The 4-point and 5-point conformal blocks exist only in the comb channel. The 6-point conformal blocks can also exist  in the snowflake channel.} This is due to the fact that the procedure of constructing a geodesic Witten diagram requires  vertices to have only an  even number of bulk-to-boundary propagators, whereas a Witten diagram with an odd number of boundary points does not meet this requirement.

\section{AdS integral identities and modified propagators }
\label{sec:new}

The decomposition method described in the previous section  applies to  Witten diagrams, i.e. to AdS Feynman diagrams   with boundary endpoints only. In order to have an expansion into single-trace and double-trace terms for AdS Feynman diagrams in the bulk (i.e. such functions which reproduce the conformal block decomposition \eqref{4pt_expansion} when restricted on the boundary)  one needs to have a bulk version of the propagator identities of the type  \eqref{geodesic_propid1}. In this paper, we show that in the $3$-point case this can be achieved by using  three integral identities valid in Euclidean \ads space. Note that the AdS integrals  in the Poincare coordinates \eqref{metric} are given by 
\be
\label{int_poincare}
\int_{\mathbb{D}_0}d^2\bx\, \sqrt{g(\bx)} \;\;\to\;\; \int_{\mathbb{D}_1} \frac{du}{u^2} \int_{\mathbb{D}_2}dz\,,
\ee
where $\mathbb{D}_{0,1,2}$ are some integration domains.  The $z$-variable in the integral identities given below is extended to the complex plane,  yielding particular integration contours that encircle the resulting  poles of the integrands. Depending on whether $\mathbb{D}_0$ has compact dimensions, the domains $\mathbb{D}_{1,2}$ can be compact or not.  

\subsection{Three  identities}

\begin{figure}
\centering
\includegraphics[scale=0.9]{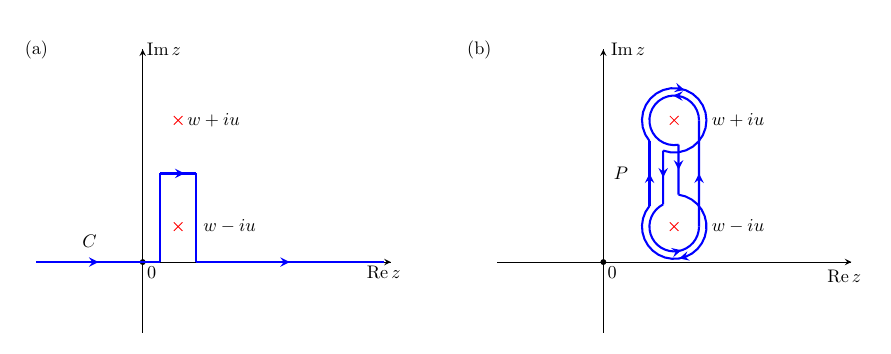}
\caption{{\bf (a)} The contour $C$  on the complex $z$-plane $\mathbb{C}$. The red crosses are poles of the analytically continued propagator $\widehat{G}_h(\bx(u,z),\bx'(u',z'),w)$.  {\bf (b)} The Pochhammer contour $P$ in $\mathbb{C}\symbol{92} \{w-iu,w+iu\}$. }
\label{fig:poch}
\end{figure}

\begin{itemize}

\item  The conversion identity: 
\be 
\label{intro_simpleprop}
G_h(\bx,\bx') = \int_{z'-iu'}^{z'+iu'}d w\;\widehat{G}_h(\bx,\bx',w)\,.
\ee
\item The superposition  identity: 
\be 
\label{intro_propid}
\ba{r}
\dps
\int_{\text{AdS}_2}d^2\bx \sqrt{g(\bx)}\;G_h(\bx,\bx')f(\bx)= \int_{z'-iu'}^{z'+iu'}d w \int_{0}^{\infty} \frac{du}{u^2}
\int_{C}dz\;\widehat{G}_h(\bx,\bx',w)f(\bx) \, +
\vspace{2.5mm}  
\\
\dps
+\,\pi
\int_{0}^{u'} \frac{du}{u^2}\int_{z'-i(u-u')}^{z'+i(u-u')}dz\;\widetilde{G}_h(\bx,\bx')f(\bx)\,.
\ea
\ee
\item The splitting  identity:
\be 
\label{intro_double-trace}
\ba{l}
\dps
\int_{0}^{u_3} \frac{du}{u^2}\int_{z_3-i(u-u_3)}^{z_3+i(u-u_3)}dz\;G_{\h_1}(\bx,\bx_1)G_{\h_2}(\bx,\bx_2)\widetilde{G}_{h_3}(\bx,\bx_3) = 
\vspace{3mm}  
\\
\dps
= \sum_{n=0}^{\infty} \frac{
a(\h_3;\h_1,\h_2;n)}{\rho_{h_1, h_2, n}}\, \int_{z_3-iu_3}^{z_3+iu_3}d w\;\oint_{0} \frac{du}{u^2} \oint_{P} dz\, 
G_{\h_1}(\bx,\bx_1)G_{\h_2}(\bx,\bx_2) \widehat{G}_{\dth_{12|n}}(\bx,\bx_3,w)\,. 
\ea
\ee
\end{itemize}
Here, the bulk coordinates are $\bx = (u,z)$, $\bx' = (u', z')$,  $f(\bx)$ is a test function,  $\rho_{h_1, h_2, n}$ is some coefficient, and 
\be 
\label{propagators}
\ba{l}
\widehat{G}_h(\bx,\bx',w) = K_h(\bx, w)\mathbb{K}_{\h}(\bx', w)\,,
\vspace{5mm}  
\\
\dps
\widetilde{G}_h(\bx,\bx') = \frac{-2i}{4^{\h}}\frac{\Gamma(2\h)}{\Gamma(\h)\Gamma(\h)}\left(\frac{\Gamma(1-2h)}{\Gamma(1-h)^2} G_h(\bx,\bx') + \frac{\Gamma(2h-1)}{\Gamma(h)^2} G_{1-h}(\bx,\bx')\right).
\ea
\ee 
Functions $\widehat{G}_h$ and $\widetilde{G}_h$  were originally introduced in \cite{Kabat:2011rz} as propagators in  Lorentzian \ads space  (see our discussion in section \bref{sec:comment}). They are solutions of  the homogeneous Klein-Gordon (KG) equation. Note that $\widehat{G}_h = \widehat{G}_h(\bx(u,z),\bx'(u',z'),w)$ arising in  \eqref{intro_simpleprop}--\eqref{intro_double-trace} is  analytically continued on the complex $z$-plane, where it has poles at $z = w \pm iu$,  $w\in \CC$, $u\in \RR$. The corresponding integration contours $C$ and $P$ on the complex plane   are shown in fig. \bref{fig:poch}. A detailed discussion of the above  identities will be given in section \bref{sec:prop}.

\subsection{Modified AdS Feynman diagrams}
\label{sec:mod}

Using  the two modified propagators \eqref{propagators} instead of the standard ones results in   modified AdS Feynman diagrammatics. Several useful properties  highlight  different roles of $\widehat{G}$ and $\widetilde{G}$. Simultaneously,  the integral identities allow one to decompose  standard  AdS Feynman diagrams into infinite sums of modified AdS Feynman diagrams.  

\paragraph{Decomposition.} Let us consider the $3$-point AdS Feynman diagram, which we schematically write as   $\int GGG$. One applies the identities \eqref{intro_simpleprop}--\eqref{intro_double-trace}  in four steps. 

\begin{enumerate}

\item Apply the superposition identity \eqref{intro_propid} to any $G$ inside the AdS Feynman  diagram. The result is the sum of two terms: $\int GGG \to \int \widehat{G}GG + \int \widetilde{G}GG$.

\item Repeat this step in the first term $\int \widehat{G}GG$  for another $G$ until no $G$ are left. The term without $\widetilde{G}$ is the single-trace contribution.

\item Replace each $G$ in the every term by $\widehat{G}$ using the conversion identity \eqref{intro_simpleprop}.

\item Convert each term with  $\widetilde{G}$ into the term with  $\widehat{G}$ by means of  the splitting identity \eqref{intro_double-trace}. This produces an infinite sum of double-trace contributions.

\end{enumerate}

\noindent Here and below, the terminology  {\it single-trace} and {\it double-trace} is used in the following sense.  A contribution is considered single-trace if the modified propagators in an AdS Feynman diagram have weights that are not composites of the original $h_i$ ($i=1,2,3$). Conversely, a contribution is considered  as double-trace if at least one of the modified propagators has a composite weight   $h_{ij|n} = h_i+h_j+2n$ ($n=0,1,2,...$). 

Schematically, the above  algorithm can be represented as follows:  
\be 
\label{scheme}
\ba{c}
\dps
\int G_1 G_2 G_3 \;\stackrel{\eqref{intro_propid}}{\longrightarrow}\; 
\int {\color{blue} \widehat{G}_1} G_2 G_3 + \int {\color{purple} \widetilde{G}_1} G_2 G_3 
\;\stackrel{\eqref{intro_propid}}{\longrightarrow} \int {\color{blue} \widehat{G}_1 \widehat{G}_2} G_3 + \int {\color{purple} \widetilde{G}_1} G_2 G_3+ \int {\color{blue} \widehat{G}_1} {\color{purple} \widetilde{G}_2} G_3
\vspace{3mm}  
\\
\dps
\stackrel{\eqref{intro_propid}, \,\eqref{intro_simpleprop}}{\longrightarrow} 
\int {\color{blue} \widehat{G}_1 \widehat{G}_2 \widehat{G}_3} 
+ \int {\color{purple} \widetilde{G}_1} {\color{blue} \widehat{G}_2 \widehat{G}_3} 
+ \int  {\color{blue} \widehat{G}_1 } {\color{purple} \widetilde{G}_2}  {\color{blue} \widehat{G}_3} 
+ \int  {\color{blue} \widehat{G}_1 \widehat{G}_2} {\color{purple} \widetilde{G}_3}
\ea
\ee
$$
\ba{l}
\dps
\stackrel{\eqref{intro_double-trace}}{\longrightarrow} \underbrace{\int {\color{blue} \widehat{G}_1 \widehat{G}_2 \widehat{G}_3} }_{\text{single-trace}}\;\;
 +\;\; \Big(\underbrace{\sum_{n}a^{23}_n \int  {\color{blue} \widehat{G}_{23|n} \widehat{G}_2 \widehat{G}_3} + (1 \leftrightarrow 2)+ (1 \leftrightarrow 3)}_{\text{double-trace}} \Big),
\ea
$$ 
where $a^{23}_n$ are fixed coefficients and  $G_i = G_{h_i}(\bx,\bx_i)$, $G_{ij|n} = G_{\dth_{ij|n}}(\bx,\bx_k)$, $ i\neq j\neq k\neq i$. 

Note that we obtained the second line in  \eqref{scheme} only  by means of the conversion and superposition identities:
\be 
\label{half_expansion}
\int G_1 G_2 G_3  = 
\int { \widehat{G}_1 \widehat{G}_2 \widehat{G}_3} 
+ \int { \widetilde{G}_1} {\widehat{G}_2 \widehat{G}_3} 
+ \int  {\widehat{G}_1 } { \widetilde{G}_2}  { \widehat{G}_3} 
+ \int  {\widehat{G}_1 \widehat{G}_2} { \widetilde{G}_3}\,.
\ee

\paragraph{Equations of motion.} Let us act with the KG operators $\Box_{\bx_i} - h_i (h_i-1)$, where $\Box_{\bx_i}$ is the Laplace-Beltrami operator in local coordinates $\bx_i\in$AdS$_2$, to   each term on the right-hand side of the decomposition  \eqref{half_expansion}:

\be
\label{eom_modified}
\hspace{-74mm}(\Box_{\bx_i} - h_i (h_i-1)) \int {\widehat{G}_1 \widehat{G}_2 \widehat{G}_3} = 0\,,\quad i = 1,2,3\,;
\ee
$$
(\Box_{\bx_l} - h_l (h_l-1))\int {\widetilde{G}_i} \widehat{G}_j \widehat{G}_k = - \delta_{il} \frac{2\sqrt{\pi}\,\Gamma(h_i+\half)}{\Gamma(h_i)}\,G_{h_j}(\bx_j,\bx_i)G_{h_k}(\bx_k,\bx_i)\,,\quad i \neq j\neq k \neq i\,,
$$
where $\delta_{il}$ is the Kronecker symbol. We conclude  that  terms that contain  the propagator $\widetilde{G}_{h_i}(\bx,\bx_i)$ satisfy  the inhomogeneous KG equation in $\bx_i$ and the homogeneous KG equation in $\bx_j \neq \bx_i$, whereas  terms which do not contain this propagator satisfy  the homogeneous KG equations. In other words, the presence or  absence of $\widetilde{G}_{h_i}(\bx,\bx_i)$  completely defines how the KG operators act on the modified AdS Feynman diagram. Similarly, the modified AdS Feynman diagram  containing  the propagator $\widehat{G}_{h_i}(\bx,\bx_i,w_i)$ satisfies  the homogeneous KG equation in $\bx_i$.

\paragraph{Boundary asymptotics.}  By direct calculation one shows that the leading asymptotics ($u_i\to 0$) in the decomposition  \eqref{half_expansion} are given by 
\be 
\label{prop_boundary}
\ba{l}
\dps
\hspace{-27mm}\int { \widehat{G}_1 \widehat{G}_2 \widehat{G}_3}  = \prod_{k = 1}^3  \int_{z_k+iu_k}^{z_k-iu_k}dw_k\;\int_0^{\infty}\frac{du}{u^2}\int_{C} dz\;\prod_{m = 1}^3\widehat{G}_{h_m}(\bx,\bx_m,w_m)  
\vspace{3mm}
\\
\dps
\hspace{-12mm}\stackrel{u_{1,2,3}\to0}{\simeq } u_1^{h_1} u_2^{h_2} u_3^{h_3}\int_0^{\infty}\frac{du}{u^2}\int_{\RR} dz\; K_{\h_1}(\bx,z_1) K_{\h_2}(\bx,z_2) K_{\h_3}(\bx,z_3)\,,
\vspace{3mm}
\ea 
\ee
$$
\ba{l}
\dps
\int { \widetilde{G}_i} {\widehat{G}_j \widehat{G}_k}  =  \prod_{m = j,k}\int_{z_m+iu_m}^{z_m-iu_m}dw_m\int_{0}^{u_i} \frac{du}{u^2}\int_{z_i-i(u-u_i)}^{z_i+i(u-u_i)} \hspace{-2mm}dz\;
\widetilde{G}_{h_i}(\bx,\bx_i) \widehat{G}_{h_j}(\bx,\bx_j,w_j)  \widehat{G}_{h_k}(\bx,\bx_k,w_k)  
\vspace{3mm}
\\
\dps
\hspace{15mm}\stackrel{u_{1,2,3}\to0}{\simeq } \eta\, u_i^{h_j+h_k} u_j^{h_j} u_k^{h_k} \oint_{0} \frac{du}{u^2} \oint_{P[z_i-iu,z_i+iu]} \hspace{-2mm} dz\, 
K_{\h_j+\h_k}(\bx,z_i) K_{\h_j}(\bx,z_j) K_{\h_k}(\bx,z_k) \,,
\ea
$$
where $\eta$ is some coefficient, the contour $C$ is shown in fig. \bref{fig:contour_KKK}. The mnemonic rule here is that, near the boundary, the bulk-to-boundary propagator $u_i^{h_i}K_{h_i}(\bx,z_i)$ \eqref{bulk-to-boundary} corresponds to the modified propagator $\widehat{G}_{h_i}(\bx,\bx_i,w_i)$, whereas  the propagator $u_i^{h_j+h_k}K_{h_j+h_k}(\bx,z_i)$ corresponds to the modified propagator $\widetilde{G}_{h_i}(\bx,\bx_i)$, with  $i\neq j \neq k\neq i$. This  correspondence is valid  only in the context of the modified $3$-point AdS Feynman diagrams,  because the near-boundary behavior of standard and modified propagators differs.  Specifically,  the integration domains of modified diagrams depend on $u_i$, which directly modifies their asymptotic behavior near the boundary. 

Now, substituting the above asymptotic relations into \eqref{half_expansion} one obtains (schematically) 
\be 
\label{GGG_boundary}
\ba{l}
\dps
u_1^{h_1} u_2^{h_2} u_3^{h_3}\int K_1K_2K_3 \;\simeq \;  u_1^{h_1} u_2^{h_2} u_3^{h_3}\int  K_1 K_2 K_3 +  u_1^{h_2+h_3} u_2^{h_2} u_3^{h_3}\int K_{2+3} K_2 K_3
\vspace{3mm}
\\
\dps
\hspace{32mm} +  u_1^{h_1} u_2^{h_1+h_3} u_3^{h_3} \int  K_1 K_{1+3}  K_3 +  u_1^{h_1} u_2^{h_2} u_3^{h_1+h_2} \int  K_1 K_2 K_{1+2}\,,
\ea 
\ee 
where we simplified our notation for integrals and denoted the bulk-to-boundary propagators  as $K_i \equiv  K_{h_i}(\bx,z_i)$, $K_{j+k} \equiv  K_{h_j+h_k}(\bx,z_i)$, $i\neq j\neq k\neq i$. The first term on the right-hand side has the same leading behavior as the left-hand side. Note that the other three terms are sub-leading if the weights $h_i$ are subject to the triangle inequalities 
\be 
\label{triangle_identity}
h_1< h_2+h_3\,,\quad
h_2< h_1+h_3\,,\quad
h_3< h_1+h_2\,,
\ee 
otherwise the relation \eqref{GGG_boundary} is incorrect. This reflects the fact that the $3$-point Witten diagram on the left-hand side of \eqref{GGG_boundary} diverges when the triangle inequalities are not satisfied. The  three sub-leading terms come from the modified diagrams containing  $\widetilde{G}_{h_i}$, which correspond to the double-trace terms according to the decomposition \eqref{scheme}. In other words, the double-trace terms near the boundary fall off faster than the single-trace term provided that the triangle inequalities are satisfied. 

\paragraph{Contact diagrams.} As the superposition identity \eqref{intro_propid} is formulated with a test function $f(\bx)$, one can expand the $n$-point {\it contact} AdS Feynman diagram into modified diagrams by choosing a  test function  to be a product of $n-1$ bulk-to-bulk propagators. Then, applying the superposition identity $n$ times one obtains 
\be 
\label{half_expansion_n}
\int \prod_{i=1}^n G_i = 
\int \prod_{i=1}^n {\widehat{G}_i} 
+ \int {\widetilde{G}_1} \prod_{i=2}^n{\widehat{G}_i} 
+ \int  { \widetilde{G}_2}\prod_{\substack{i=1\\i\neq 2}}^n {\widehat{G}_i}  
+...+ \int  {\widetilde{G}_n} \prod_{i=1}^{n-1} { \widehat{G}_i} \,,
\ee
where we also used the conversion identity \eqref{intro_simpleprop} to replace  every $G$ on the right-hand side by $\widehat G$. Note that the action of the KG operators on  \eqref{half_expansion_n} yields a set of relations similar to \eqref{eom_modified}. The roles of the modified  propagators $\widetilde{G}$ and $\widehat{G}$ in the $n$-point case remain the same. 

The near-boundary ($u_i \to 0$) asymptotic expansion of \eqref{half_expansion_n}  is given by
\be 
\label{Gn_boundary}
\prod_{i=1}^n u_i^{h_i} \int \prod_{j=1}^nK_j \;\simeq \;  \prod_{i=1}^n u_i^{h_i} \int \prod_{j=1}^nK_j +  \sum_{k=1}^n \prod_{\substack{i=1\\i\neq k}}^n (u_k u_i)^{h_i}\int K_{1+...+\cancel{k}+...+n} \prod_{\substack{j=1\\j\neq k}}^nK_j\,,
\ee 
where we defined $K_{1+...+\cancel{k}+...+n} \equiv K_{h_1+...+\cancel{h}_k+...+h_n}(\bx,z_k)$ and taken only the leading asymptotics of each term. The analysis of  \eqref{Gn_boundary} is essentially the same as in the 3-point case, the only difference is that in the $n$-point case the propagator $\widetilde{G}_{h_k}(\bx,\bx_k)$ corresponds to the bulk-to-boundary propagator $u_k^{h_1+...+\cancel{h}_k+...+h_n}K_{1+...+\cancel{k}+...+n}$. Also note that the first term in  \eqref{Gn_boundary} is leading only if the weights are subject to the  generalized triangle inequalities
\be 
\label{triangle_identity1}
h_1+...-h_k+...+h_n \geq 0 \,,\qquad \forall k\in \{1,...,n\}\,.
\ee

\subsection{Commentary}
\label{sec:comment}
Firstly, we note that  the  identities  \eqref{intro_simpleprop} and  \eqref{intro_propid} have much in common with an identity  relating  the  bulk-to-bulk and modified propagators defined in  Lorentzian AdS$_2$ space \cite{Kabat:2011rz}:
\be 
\label{intro_Kabat_rel}
G_h(\bx,\bx') = \pi\, \theta(u'-u)\theta(u'-u-|z'-z|)\,\widetilde{G}_h(\bx,\bx') + i\int_{z'-u'}^{z'+u'}dw\;\widehat{G}_h(\bx,\bx',w)\,,
\ee 
where $\theta(x)$ is the Heaviside step function. Consider  the conversion identity \eqref{intro_simpleprop}. In a domain where the step functions vanish,   \eqref{intro_Kabat_rel} is a special case of \eqref{intro_simpleprop}. On the other hand, the identity \eqref{intro_Kabat_rel} can be integrated  over AdS$_2$ space using a test function  and subsequently subjected to the Wick rotation. One can show that the result is given by the superposition identity \eqref{intro_propid}, provided a specific class of test functions is chosen. See our discussion around \eqref{Kabat_rel} in Appendix \bref{app:lemma4}. 

Secondly, the superposition  identity \eqref{intro_propid} can be applied to any AdS Feynman diagram; specifically,  a test function $f(\bx)$ can be chosen as a product of any number of propagators. On the other hand, the splitting  identity \eqref{intro_double-trace} transforms the product of only three bulk-to-bulk  propagators, therefore, it can be applied  only when an AdS Feynman diagram has at least one $3$-valent vertex. The splitting identity can be seen as a direct analogue of the propagator identity \eqref{geodesic_propid1} in the sense that both identities produce  infinite sums of  double-trace terms and add one-dimensional integration (i.e. a number of integrations on the left and right differs by one). 

Thirdly, we note that the single-trace contribution is directly obtained by replacing each  bulk-to-bulk propagators $G$ by the modified propagator $\widehat G$, see the last line in \eqref{scheme}. In the context of $2d$ thermal conformal correlators this trick was proposed in \cite{Kraus:2017ezw}. The authors changed the bulk-to-bulk propagator inside the $2$-point Witten diagram in the thermal AdS$_3$ as 
\be 
\label{Kraus_change}
G_h(x_1,x_2) \longrightarrow \int_{\partial \text{AdS}_3} dz d \bar{z} \;\widehat{G}_h(x_1,x_2; z, \bar{z})\,,
\ee 
where the conformal boundary $\partial \text{AdS}_3\cong \mathbb{T}^2$. The integrand is the AdS$_3$ version of the modified propagator $\widehat{G}_h(\bx,\bx',w)$  \eqref{propagators}. This propagator  contains the Heaviside step functions restricting  the integration domain like in the standard formulation of the HKLL reconstruction (see the discussion below \eqref{smear}). The substitution \eqref{Kraus_change} results in the modified Witten diagram which obeys the Casimir equation for the 2-point torus conformal block \cite{Kraus:2017ezw}.

\section{Wilson network decomposition of AdS Feynman diagrams}
\label{sec:prop}

The purpose of this section is twofold. First, we rigorously   derive the propagator identities  discussed in the previous section. Second, we show that their application yields an infinite expansion of the 3-point \ads Feynman diagram with bulk endpoints, wherein  each term  corresponds to the 3-point Wilson line network in \ads space carrying the particular conformal  weights.

\subsection{Wilson networks as basis functions}
\label{sec:wilson_basis}

Let us briefly recap what we mean by the Wilson line networks in \ads space, which has an isometry algebra  $\sltwo$ \cite{Alkalaev:2023axo} (see also  \cite{Bhatta:2016hpz,Besken:2016ooo,Bhatta:2018gjb,Castro:2018srf,Alkalaev:2020yvq}). The $3$-point Wilson line network is built by contracting  $3$-valent $\sltwo$ intertwiner carrying  external weights $h_1,h_2,h_3$ with three Wilson line operators of weights $h_1,h_2, h_3$ connecting  the point $\bx = 0$ and the points $\bx_1,\bx_2, \bx_3$ in the bulk. Thus, this is an operator in the tensor product of three (infinite-dimensional) $\sltwo$ Verma modules. Let $\cV_{h_1h_2h_3}(\bx_1,\bx_2, \bx_3)$  be a particular matrix element of the Wilson line network, it is evaluated  between three cap states which are the Ishibashi states. In what follows, we refer to these matrix elements  as $3$-point {\it AdS vertex functions}. The AdS vertex functions can be explicitly calculated as multivariate generalized hypergeometric series  \cite{Alkalaev:2024cje}. The $n$-point construction can be found in  \cite{Alkalaev:2023axo}. 

The crucial property of the $n$-point AdS vertex functions is that they are HKLL reconstructed $n$-point global conformal blocks in the comb channel \cite{Alkalaev:2024cje} (see \eqref{3ptVertex} below). Here, we  keep the leading contribution of the $1/N$ expansion, where the role of $N$ is played by the large central charge. Therefore, the Virasoro conformal blocks with light operators ($h = O(c^0)$) in the limit $c\to \infty$ reduce to the global conformal blocks, which are $\sltwo$ conformal blocks in this case.

The $3$-point AdS Feynman diagram in the bulk can be expanded into the $3$-point AdS vertex functions \cite{Alkalaev:2024cje}:\footnote{The $3$-point AdS Feynman diagram with two boundary points was calculated in \cite{Zhou:2018sfz}, and then expressed in terms of the $3$-point geodesic Witten diagrams in \cite{Giombi:2020xah}.  On the other hand, the 3-point AdS vertex function with two boundary points reproduces the 3-point geodesic Witten diagram \cite{Alkalaev:2024cje}. An explicit expression for the $3$-point AdS Feynman diagram with three bulk points was found in \cite{Jepsen:2019svc}.} 
\be 
\label{GGG_rel}
\hspace{-20mm}
\int_{\text{AdS}_2}d^2\bx\sqrt{g(\bx)}\, G_{\h_1}(\bx,\bx_1) G_{\h_2}(\bx,\bx_2)G_{\h_3}(\bx,\bx_3) 
\ee
$$
\ba{l}
\dps
= \coef_0\cV_{\h_1\h_2\h_3}(\bx_1,\bx_2,\bx_3)
+\sum_{n=0}^{\infty}\coef^{^{(1|23)}}_n\cV_{\dth_{23|n} \h_2\h_3}(\bx_1,\bx_2,\bx_3)
\vspace{3mm}
\\
\dps
+\sum_{n=0}^{\infty}\coef^{^{(2|13)}}_n\cV_{\h_1 \dth_{13|n}\h_3}(\bx_1,\bx_2,\bx_3)
+\sum_{n=0}^{\infty}\coef^{^{(3|12)}}_n\cV_{\h_1 \h_2\dth_{12|n}}(\bx_1,\bx_2,\bx_3)\,,
\ea
$$ 
where the $\coef$-coefficients are given by 
\be 
\label{d_coef}
\ba{l}
\dps 
\coef_0 =  \frac{\pi^{\half}}{2}\frac{\Gamma\big(\frac{\h_1+\h_2+\h_3}{2}-\half\big)}{\pref_{\h_1 \h_2\h_3}}\, \frac{\Gamma(\frac{-\h_1+\h_2+\h_3}{2})\Gamma(\frac{\h_1-\h_2+\h_3}{2})\Gamma(\frac{\h_1+\h_2-\h_3}{2})}{\Gamma(\h_1)\Gamma(\h_2)\Gamma(\h_3)}\,,
\vspace{3mm}
\\
\dps
\coef^{^{(i|jk)}}_n = \frac{a(\h_i;\h_j,\h_k;n)}{\pref_{\h_j \h_k\dth_{jk|n}}}\,,
\ea
\ee 
where the $a$-coefficients are given by  \eqref{a_coef} (see our comment below \eqref{coef_4pt}),  the  $3$-point structure constants are given by  
\be
\label{prefactor}
\pref_{\h_1\h_2\h_3} = \left[\frac{(-2\h_1)!(-2\h_2)!(-2\h_3)!}{\Delta(\h_1,\h_2,\h_3)}\right]^{\half}\,,
\ee
where $\Delta(a,b,c) = (-a-b-c+1)!(c-a-b)!(b-a-c)!(a-b-c)!$ is the modified triangle function. 
A diagrammatic representation of this expansion is shown in fig. \bref{fig:GGG_as_V}.

AdS vertex functions in the Wilson network decomposition \eqref{GGG_rel} play the same role as conformal blocks in the conformal block decomposition of the $4$-point exchange Witten diagram \eqref{4pt_expansion}, in particular, the  groups of single-trace and double-trace terms in the two decompositions  have the same conformal dimensions.  In other words, the relation \eqref{GGG_rel}, which was found by brute force  calculation of the AdS Feynman diagram,  hints at possibility of studying $n$-point AdS Feynman diagrams by expanding them into $n$-point AdS vertex functions, which, therefore, can be treated as a sort of basis functions in the space of  bulk correlation functions. (Similarly, the conformal blocks are basis functions in the space of  conformal correlation functions.) At the boundary,  they are  directly  reduced to conformal blocks of single-trace and double-trace intermediate dimensions (one-to-one, without any further transformations or resummations).  This provides the conformal block expansion of   Witten diagrams. In our $n=3$ case the  decomposition \eqref{GGG_rel} on the conformal boundary reduces  to the single-trace term, which is the  $3$-point conformal correlation function of weights $h_1, h_2, h_3$ (this is  the first term on the right-hand side of \eqref{GGG_rel}).\footnote{This happens only when  the triangle inequalities \eqref{triangle_identity} are satisfied. In this case the single-trace term is leading near the boundary, whereas  the double-trace terms are sub-leading \cite{Alkalaev:2024cje} (see also our discussion of the boundary asymptotics in section \bref{sec:mod}).  If the triangle inequalities are not satisfied, then the $3$-point AdS Feynman diagram on the left-hand side of \eqref{GGG_rel} diverges that requires a regularization, see \cite{Castro:2024cmf} for details.}

\subsection{Deriving the Wilson network decomposition}
\label{sec:explcit}

The form of  propagator identities \eqref{intro_simpleprop}--\eqref{intro_double-trace} is very specific, so it is not obvious at all how to find them or other identities of the same type (if any). Consequently, we derive a convenient integral representation of the AdS vertex function that implicitly encodes all three identities and, eventually, the Wilson network decomposition \eqref{GGG_rel}. 

\subsubsection{Single-trace contributions}
\label{sec:single}

Consider the holographic reconstruction formula for  the $3$-point AdS vertex function \cite{Alkalaev:2024cje}
\be 
\label{3ptVertex}
\ba{l}
\dps
\hspace{-2mm}\cV_{\h_1\h_2\h_3}(\bx_1,\bx_2,\bx_3)=
  \prod_{k = 1}^3 \int_{z_k-iu_k}^{z_k+iu_k}dw_k\;\mathbb{K}_{h_k}(\bx_k,w_k)\,\big\langle\cO_{\h_1}(w_1)\cO_{\h_2}(w_2)\cO_{\h_3}(w_3)\big\rangle\,,
\ea
\ee
where $\mathbb{K}_{h_k}(\bx_k,w_k)$ is the HKLL smearing function \eqref{smear}, the $3$-point conformal correlation function  is given by
\be
\label{3pt_conf}
\braket{\cO_{\h_1}(w_1)\cO_{\h_2}(w_2)\cO_{\h_3}(w_3)} =  \frac{\pref_{\h_1\h_2\h_3}}{
(w_1-w_2)^{\h_1+\h_2-\h_3}\,(w_2-w_3)^{\h_2+\h_3-\h_1}(w_3-w_1)^{\h_1+\h_3-\h_2}}\,,
\ee
with  the $\pref$-coefficient \eqref{prefactor}, the integral over $w_k$ is a complex line integral over the line segment connecting points $z_k-iu_k$ and $z_k+iu_k$. The integrals in \eqref{3ptVertex} converge only for $h_i\geq0$. The analytic continuation of this relation for $h_i<0$ can be obtained by changing integration domains from the line segments to the Pochhammer contours, see \cite{Alkalaev:2024cje} for details. Moreover, as we show below, further restriction $h_i> \half$ is necessary for our analysis, which we will assume from now on.

Note that the intertwiners in the Wilson network construction which define the left-hand side of \eqref{3ptVertex} constrain the weights such that they satisfy the selection rule $\h_2+\h_3-\h_1\in\ZZ$. However, the right-hand side of  \eqref{3ptVertex} can be used to define the AdS vertex functions for arbitrary non-negative weights $h_i$. From now on, we understand  AdS vertex functions as the HKLL reconstruction of boundary conformal blocks.

If $w_i\in\mathbb{R}$, then the 3-point function \eqref{3pt_conf}  has known AdS integral representation in terms of the $3$-point Witten diagram \cite{Freedman:1998tz}
\be 
\label{KKK}
\big\langle{\cO_{\h_1}(w_1)\cO_{\h_2}(w_2)\cO_{\h_3}(w_3)}\big\rangle  = \frac{1}{\coef_0} \int_{\text{AdS}_2}d^2\bx\sqrt{g(\bx)}\;K_{h_1}(\bx,w_1)K_{\h_2}(\bx,w_2)K_{\h_3}(\bx,w_3)\,,
\ee 
where the $\coef_0$-coefficient is given by \eqref{d_coef}, the conformal dimensions $h_i>\half$ are subject to the triangle inequalities \eqref{triangle_identity}. The conditions above, including reality of $w_i$, are necessary for the convergence of the AdS integral \eqref{KKK}.

If $w_i \in \CC$, then the right-hand side of \eqref{KKK} diverges. To see this, rewrite the AdS$_2$ integral as a double integral over $u$ and $z$ \eqref{int_poincare}   and consider poles of the integrand located at $w_i\pm iu$ on the complex $z$-plane, $i = 1,2,3$, which  directly follow from the form of the bulk-to-boundary propagators \eqref{bulk-to-boundary}. If the integration variable $u$ takes values $u = |\im(w_i)| \neq 0$, then the integration contour on the $z$-plane crosses one of the poles $w_i+ iu$ or $w_i- iu$ that makes the integral \eqref{KKK} divergent. 

As the 3-point conformal correlator in the reconstruction formula \eqref{3ptVertex} is  taken on the complex plane,  one should first analytically continue the integral representation \eqref{KKK} to the complex plane by deforming the integration contour in the complex  $z$-plane to avoid crossing the poles. One can show that the analytically continued function  reads 
\be 
\label{analytic_cont_KKK}
\ba{l} 
\dps 
\frac{\coef_0\,\pref_{\h_1 \h_2\h_3}}{
(w_1-w_2)^{\h_1+\h_2-\h_3}\,(w_2-w_3)^{-\h_1+\h_2+\h_3}(w_3-w_1)^{\h_1-\h_2+\h_3}} =
\vspace{3mm}  
\\
\dps
\hspace{40mm}= \int_0^{\infty}\frac{du}{u^2}\int_Cdz\;K_{\h_1}(\bx,w_1)K_{\h_2}(\bx,w_2)K_{\h_3}(\bx,w_3)\,,
\ea 
\ee 
where the $\coef_0$-coefficient is given by \eqref{d_coef}  and $C$ is the  contour shown in fig. \bref{fig:contour_KKK} (see Appendix \bref{app:Witten_analytic} for the proof).
\begin{figure}
\centering
\includegraphics[scale=0.9]{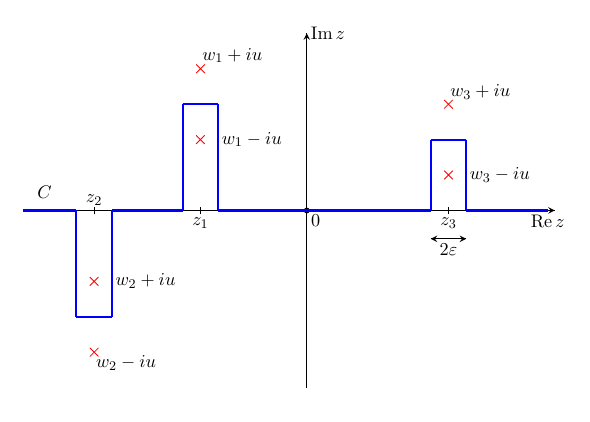}
\caption{The contour $C$ on the complex $z$-plane. The red crosses are poles of the integrand \eqref{3ptVertexInterm}; the regularization parameter  $\varepsilon \to 0$.}
\label{fig:contour_KKK}
\end{figure}
Substituting \eqref{analytic_cont_KKK} back into the  reconstruction formula \eqref{3ptVertex} one obtains
\be 
\label{3ptVertexInterm}
\ba{l}
\dps
\cV_{\h_1\h_2\h_3}(\bx_1,\bx_2,\bx_3)=
\frac{1}{\coef_0}  \prod_{k = 1}^3  \int_{z_k-iu_k}^{z_k+iu_k}dw_k\;\int_0^{\infty}\frac{du}{u^2}\int_Cdz\;\prod_{m = 1}^3\widehat{G}_{h_m}(\bx,\bx_m,w_m)\,,
\ea
\ee
where the modified propagators are defined in  \eqref{propagators}. 

Recalling that the single-trace term in the expansion \eqref{GGG_rel} is given by the AdS vertex function we see that the integral formula \eqref{3ptVertexInterm} can be obtained from the 3-point AdS Feynman  diagram by replacing the bulk-to-bulk propagators $G$ with the modified propagators $\widehat{G}$, cf. \eqref{scheme}. However,  $\widehat{G}$ has   an additional complex variable which should be integrated out by adding one more integral  along with changing  the integration contours  to avoid poles of the new integrand.

\subsubsection{Double-trace contributions}
\label{sec:double}

To find a similar operation that extracts double-trace contributions, one considers the integral  representation \eqref{3ptVertexInterm} and divides the integration contour $C$ into four  parts: the first part $C_{0}(\varepsilon)$ contains every real line segment of $C$, the other three parts are $\Pi$-shaped  contours bypassing the  points  $z_1, z_2, z_3$ on the real axis $\re z$, see fig. \bref{fig:contour_KKK}:
\be
\label{split_C}
C = C_{0}(\varepsilon) \cup C_1(\varepsilon)\cup C_2(\varepsilon)\cup C_3(\varepsilon) \,,
\ee
where $\varepsilon$ denotes a half-width  of each $\Pi$-shaped part $C_{1,2,3}(\varepsilon)$. All segments are defined so that they are closed intervals near the poles  and we will control that the integrals under consideration  do not diverge on their endpoints. Roughly speaking, such a  splitting of the original contour underlies a decomposition of the AdS Feynman diagram into Wilson networks, where the first non-compact contour corresponds to the single-trace contribution, whereas the other three compact contours correspond to the double-trace contributions. The contour  \eqref{split_C} splits  the AdS vertex function  \eqref{3ptVertexInterm} into a sum of four integrals:
\be
\label{A_dec}
\ba{l}
\dps
\cV_{\h_1\h_2\h_3}(\bx_1,\bx_2,\bx_3) =  \sum_{i=0}^3 A^{^{(i)}}_{\h_1\h_2\h_3}(\bx_1,\bx_2,\bx_3|\varepsilon) = \sum_{i=0}^3 A^{^{(i)}}_{\h_1\h_2\h_3}(\bx_1,\bx_2,\bx_3) + O(\varepsilon)\,, 
\ea
\ee  
where each term $A^{^{(i)}}_{\h_1\h_2\h_3}(\bx_1,\bx_2,\bx_3|\varepsilon)$ is given by the integral \eqref{3ptVertexInterm} restricted on the corresponding part of the whole contour $C$, and 
\be
\label{A_limit}
A^{^{(i)}}_{\h_1\h_2\h_3}(\bx_1,\bx_2,\bx_3) \,\equiv\,  \lim_{\varepsilon \to0}A^{^{(i)}}_{\h_1\h_2\h_3}(\bx_1,\bx_2,\bx_3|\varepsilon)\,.
\ee 
At this stage, we assume that the limits in \eqref{A_limit} do exist that we explicitly verify later in this section.  By construction, the left-hand side of \eqref{A_dec} is independent of the regularization parameter $\varepsilon$. It follows that $O(\varepsilon)$ terms in \eqref{A_dec} add up to zero and, therefore, can be neglected. 
 
Let us consider the first term in \eqref{A_dec}:  
\be 
\label{first_term}
\ba{c}
\dps
A^{^{(0)}}_{\h_1\h_2\h_3}(\bx_1,\bx_2,\bx_3|\varepsilon) = \frac{1}{\coef_0}  \prod_{k = 1}^3  \int_{z_k-iu_k}^{z_k+iu_k}dw_k\;\int_0^{\infty}\frac{du}{u^2}\int_{C_0(\varepsilon)} dz\;\prod_{m = 1}^3\widehat{G}_{h_m}(\bx,\bx_m,w_m)  
\vspace{3mm}  
\\
\dps
\hspace{25mm}=\frac{1}{\coef_0}  
\int_0^{\infty}\frac{du}{u^2}\int_{C_0(\varepsilon)}dz\;\prod_{k = 1}^3 \int_{z_k-iu_k}^{z_k+iu_k}dw_k\;\widehat{G}_{h_k}(\bx,\bx_k,w_k)
\vspace{3mm}  
\\
\dps
\hspace{25mm}=\frac{1}{\coef_0} 
\int_0^{\infty}\frac{du}{u^2}\int_{C_0(\varepsilon)}dz\; G_{\h_1}(\bx,\bx_1) G_{\h_2}(\bx,\bx_2)G_{\h_3}(\bx,\bx_3)\,.
\ea
\ee
Here, the second line is obtained by interchanging the integrals over $w_k$ and $u,z$. Note that this was not possible in the integral \eqref{3ptVertexInterm} due to the fact that  $C$ depends on  $w_k$, whereas  $C_0(\varepsilon)$ does not.  The last line results from the following lemma.

\begin{figure}
\centering
\includegraphics[scale=0.9]{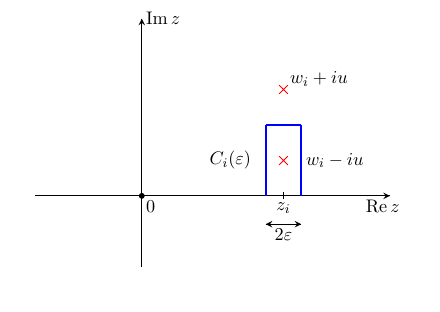}
\caption{The contour $C_i(\varepsilon)$ on the complex $z$-plane, $i=1,2,3$. The red crosses are poles in \eqref{A_def}.}
\label{fig:C_i}
\end{figure}

\begin{lemma}[the conversion identity]
The  \ads bulk-to-bulk scalar propagator has the following integral representation: 
\label{lem:PROPSIMPLE}
\be 
\label{KK_as_bb}
G_{\h}(\bx,\bx') = \int_{z'-iu'}^{z'+iu'}dw\;\widehat{G}_{h}(\bx,\bx',w)\,,
\qquad
\re(z)\neq \re(z')\;\; \text{or} \;\; u>u'-|\im(z-z')|\,, \quad h>0\,,
\ee 
where $\bx = (u,z)$, $\bx' = (u',z')$ and 
\be
\label{lem_0_G}
G_h(\bx,\bx') = \left(\frac{u u'}{u^2+u'^{\,2}+(z-z')^2}\right)^h{}_2F_1\left[\frac{\h}{2},\frac{\h}{2}+\half; \h+\half\Big|\,\left(\frac{2u u'}{u^2+u'^{\,2}+(z-z')^2}\right)^2\right],
\ee
\be
\label{lem_1_G}
\widehat{G}_h(\bx,\bx',w) = \frac{-2i}{4^{\h}}\frac{\Gamma(2\h)}{\Gamma(\h)\Gamma(\h)}\left(\frac{u'}{u'^{\,2} + (z'-w)^2}\right)^{1-\h} \left(\frac{u}{u^2 + (z-w)^2}\right)^\h\,;
\ee
$u,u' \in \RR$ and $w, z, z'\in \CC$.
\end{lemma}
The proof is given in Appendix \bref{app:lemma1}. By applying Lemma \bref{lem:PROPSIMPLE}  to  other three terms  in \eqref{A_dec} one finds that  ($i\neq j\neq k\neq i$)
\be 
\label{A_def}
\ba{l}
\dps
A^{^{(i)}}_{\h_1\h_2\h_3}(\bx_1,\bx_2,\bx_3|\varepsilon) = 
\vspace{3mm}  
\\
\dps
=\frac{1}{\coef_0} \,
\int_{z_i-iu_i}^{z_i+iu_i}dw_i\int_0^{\infty}\frac{du}{u^2}\int_{C_i(\varepsilon)}dz\;\widehat{G}_{h_i}(\bx,\bx_i,w_i) G_{\h_j}(\bx,\bx_j)G_{\h_k}(\bx,\bx_k)\,,
\ea
\ee 
where the contour  $C_i(\varepsilon)$ is shown in fig. \bref{fig:C_i}. 
Now, taking the limit $\varepsilon \to0$ in the first term yields the 3-point AdS Feynman diagram
\be
\label{A0_lim}
A^{^{(0)}}_{\h_1\h_2\h_3}(\bx_1,\bx_2,\bx_3) = 
\frac{1}{\coef_0} 
\int_{\text{AdS}_2}d^2\bx\sqrt{g(\bx)}\, G_{\h_1}(\bx,\bx_1) G_{\h_2}(\bx,\bx_2)G_{\h_3}(\bx,\bx_3)\,. 
\ee
Indeed, in this case $\lim_{\varepsilon \to 0}C_0(\varepsilon)= \RR$,  and the right-hand side in \eqref{first_term}  exactly reproduces the 3-point AdS Feynman diagram. The limit $\varepsilon \to0$ exists as the integrand has no poles in the domain $\bx\in\RR_{\geq0}\times\RR$.  

Thus, by means of \eqref{A_dec} the AdS vertex function can be represented as 
\be 
\label{vertex_as_bb}
\ba{c}
\dps
\cV_{\h_1\h_2\h_3}(\bx_1,\bx_2,\bx_3)=\frac{1}{\coef_0} 
\int_{\text{AdS}_2}d^2\bx\sqrt{g(\bx)}\, G_{\h_1}(\bx,\bx_1) G_{\h_2}(\bx,\bx_2)G_{\h_3}(\bx,\bx_3) + O(\varepsilon) 
\vspace{3mm}
\\
\hspace{15mm}
+ A^{^{(1)}}_{\h_1\h_2\h_3}(\bx_1,\bx_2,\bx_3|\varepsilon) + A^{^{(2)}}_{\h_1\h_2\h_3}(\bx_1,\bx_2,\bx_3|\varepsilon) + A^{^{(3)}}_{\h_1\h_2\h_3}(\bx_1,\bx_2,\bx_3|\varepsilon)\,.
\ea
\ee 
This formula reminds the decomposition \eqref{GGG_rel}: they both relate the $3$-point AdS Feynman  diagram and the $3$-point AdS vertex function and involve three additional (groups of) terms. Although the additional terms from these relations are different in form (integrals vs infinite sums), they are similar in their properties. E.g. these terms share the same property of transforming into each other under the cyclic permutation of conformal weights and coordinates. Also,  near-boundary ($u\to 0$) expansions of additional terms from \eqref{vertex_as_bb} and \eqref{GGG_rel} have pairwise the same powers of the $u$-coordinate.  Below we formulate three  Lemmas which are used to prove that $A^{^{(i)}}_{\h_1\h_2\h_3}(\bx_1,\bx_2,\bx_3|\varepsilon)$, $i=1,2,3$, from \eqref{vertex_as_bb} is proportional to the $(i+1)$-th term on the right-hand side of \eqref{GGG_rel} (assuming the obvious line-by-line enumeration). 
\begin{lemma} 
\label{lem:PROPTILDE}
Let $C_i(\varepsilon)$ be the contour shown in fig. \bref{fig:C_i} and $f(\bx)$ be a holomorphic function on a domain $U \times V \subset \CC^2$ of complex argument $\bx = (u,z)$, such that $U\subset \RR $ and $V \subset C_i(\varepsilon)$ for any $\varepsilon \to 0$. Let  $f(\bx)$ satisfy  $u^{-h_i}f(\bx) \to 0$ at $u \to 0$ for some $h_i > 0$. Then, the following identity holds
\be 
\label{R_def}
\ba{l}
\dps
\lim_{\varepsilon\to 0}\int_{z_i-iu_i}^{z_i+iu_i}\hspace{-2mm}dw_i\int_{0}^{\infty} \frac{du}{u^2}\int_{C_i(\varepsilon)}dz\,\widehat{G}_{h_i}(\bx,\bx_i,w_i) f(\bx) 
= \pi\int_{0}^{u_i} \frac{du}{u^2}\int_{z_i -i(u_i-u)}^{z_i +i(u_i-u)}\hspace{-1mm}dz\,\widetilde{G}_{h_i}(\bx,\bx_i)f(\bx)\,,
\ea
\ee
where $\bx_i = (u_i, z_i)$, and 
\be
\label{lem_2_G}
\widetilde{G}_{h_i}(\bx,\bx_i) = \frac{-2i}{4^{\h_i}}\frac{\Gamma(2\h_i)}{\Gamma(\h_i)\Gamma(\h_i)}\left(\frac{\Gamma(1-2h_i)}{\Gamma(1-h_i)^2} G_{h_i}(\bx,\bx_i) + \frac{\Gamma(2h_i-1)}{\Gamma(h_i)^2} G_{1-h_i}(\bx,\bx_i)\right);
\ee
$G_{h_i}(\bx,\bx_i)$ and $\widehat{G}_{h_i}(\bx,\bx_i,w_i)$ are given by \eqref{lem_0_G} and  \eqref{lem_1_G}.
\end{lemma}
The proof is given in Appendix \bref{app:lemma2}.

\begin{lemma} [the splitting identity]
\label{lem:DOUBLE-TRACE}
The following integral identity holds 
\be 
\label{G_tilde_as_vertex}
\ba{l}
\dps
\int_{0}^{u_3} \frac{du}{u^2}\int_{z_3-i(u-u_3)}^{z_3+i(u-u_3)}\hspace{-2mm}dz\;G_{\h_1}(\bx,\bx_1)G_{\h_2}(\bx,\bx_2)\widetilde{G}_{h_3}(\bx,\bx_3)= \frac{1}{8i \pi^3}\frac{(-)^{\h_1+\h_2}}{\sin(2\pi(\h_1+\h_2))}
\vspace{2.5mm}  
\\
\dps
\times \sum_{n=0}^{\infty}\frac{a(\h_3;\h_1,\h_2;n)}{\alpha'_{\h_1\h_2,n}}\,\int_{z_3-iu_3}^{z_3+iu_3} \hspace{-2mm} dw\;\oint_{0} \frac{du}{u^2} \oint_{P[w-iu,w+iu]} \hspace{-2mm} dz\, 
G_{\h_1}(\bx,\bx_1)G_{\h_2}(\bx,\bx_2) \widehat{G}_{\dth_{12|n}}(\bx,\bx_3,w)\,,
\ea
\ee 
where the $a$-coefficients are given by \eqref{a_coef}, the coefficient $\alpha'_{\h_1\h_2,n}$  is given by
\be 
\label{alpha_prime}
\alpha'_{\h_1\h_2,n} = \frac{(-)^{n}}{2\pi^{\half}i}\;\frac{\Gamma(\h_1+\h_2+n-\half)\Gamma(\h_1+n)\Gamma(\h_2+n)}{n!\,\Gamma(\h_1+\h_2+2n)\Gamma(\h_1)\Gamma(\h_2)}\,;
\ee
$P[w-iu,w+iu]$ is the Pochhammer contour around points $w-iu$ and $w+iu$, see fig. \bref{fig:poch}.
\end{lemma}
\begin{corollary}
\label{coro}
The splitting identity \eqref{G_tilde_as_vertex} can be cast into the following form:
\be 
\label{coroll_vertex}
\int_{0}^{u_3} \frac{du}{u^2} \int_{z_3-i(u-u_3)}^{z_3+i(u-u_3)}\hspace{-2mm}dz\; G_{\h_1}(\bx,\bx_1) G_{\h_2}(\bx,\bx_2) \widetilde{G}_{h_3}(\bx,\bx_3)
= \frac{1}{\pi}\sum_{n=0}^{\infty} \coef^{^{(3|12)}}_n \cV_{\h_1\h_2 \h_{12|n} }(\bx_1,\bx_2,\bx_3)\,,
\ee 
where $\cV_{\h_1\h_2 \h_{12|n} }(\bx_1,\bx_2,\bx_3)$ is the $3$-point AdS vertex function and the coefficient $\coef^{^{(3|12)}}_n$ is given by \eqref{d_coef}.
\end{corollary}
The proofs are given  in Appendix \bref{app:lemma3}. Then, it is straightforward to see that applying Lemma \bref{lem:PROPTILDE} and Lemma \bref{lem:DOUBLE-TRACE}  as well as  Corollary \bref{coro} to \eqref{A_def} one obtains
\be 
\label{double_integral}
\ba{l}
\dps
A^{^{(i)}}_{\h_1\h_2\h_3}(\bx_1,\bx_2,\bx_3)
=\frac{\pi}{\coef_0} \int_0^{u_i}\frac{du}{u^2}\int_{z_i-i(u_i-u)}^{z_i+i(u_i-u)}dz\;\widetilde{G}_{h_i}(\bx,\bx_i)G_{\h_j}(\bx,\bx_j)G_{\h_k}(\bx,\bx_k)
\vspace{3mm}  
\\
\dps
\hspace{32.5mm}= -\frac{1}{\coef_0}\,\sum_{n=0}^{\infty}\coef^{^{(i|jk)}}_n\,\cV_{\h_j \h_k \h_{jk|n}}(\bx_j,\bx_k,\bx_i)\,,
\qquad
i\neq j\neq k\neq i\,.
\ea
\ee 

Thus, along with \eqref{A0_lim} these relations  prove the decomposition  of the $3$-point AdS Feynman diagram \eqref{GGG_rel} into  $3$-point AdS vertex functions without  calculating them explicitly. Note that by acting with the KG operator $\Box_{\bx_i} - h_i (h_i-1)$ to the first and second line of \eqref{double_integral}  one obtains the following identity:
\be 
\label{eom+vertex}
\ba{c}
\dps
\frac{2\sqrt{\pi}\,\Gamma(h_i+\half)}{\Gamma(h_i)}\,G_{h_j}(\bx_j,\bx_i)G_{h_k}(\bx_k,\bx_i)
 
\vspace{3mm}  
\\
\dps
= \sum_{n=0}^{\infty}\left(h_{jk|n} (h_{jk|n}-1)-h_i (h_i-1)\right)\coef^{^{(i|jk)}}_n\,\cV_{\h_j \h_k \h_{jk|n}}(\bx_j,\bx_k,\bx_i) \,,
\qquad
i\neq j\neq k\neq i\,,
\ea
\ee 
where we  used  that AdS vertex functions solve the homogeneous KG equations \cite{Alkalaev:2023axo}.\footnote{This identity can also be  proved directly by  using explicit expressions for the 3-point AdS vertex functions found  in \cite{Alkalaev:2024cje}.}

\subsubsection{Superposition identity}

The relation \eqref{double_integral} shows that  the change  $G_{h_i} \to \widetilde G_{h_i}$, supplemented by  the change of the integration contour in the $z$-plane, allows us to  single out the double-trace terms. In other words, the relations \eqref{3ptVertexInterm} and \eqref{double_integral} present a  method of extracting  single-trace and double-trace terms from the AdS Feynman diagram by replacing bulk-to-bulk propagators with modified propagators. In what follows, we show that this can be formalized by introducing an identity which  expands a bulk-to-bulk propagator into modified propagators. 
\begin{lemma}[the superposition identity]
\label{lem:PROPID}
Let $f(\bx)$ be a test function satisfying the conditions of Lemma \bref{lem:PROPTILDE}. Let $C_i(\varepsilon)$ be the contour shown in fig. \bref{fig:C_i}, and  $C(\varepsilon)+C_i(\varepsilon)$ be a continuous curve. Here, $C(\varepsilon)$ is any contour on the complex $z$-plane that avoids  the   poles of $f(\bx)$ and satisfies $\re(z)\neq z_i$, $\forall z\in C(\varepsilon)$.  The following identity then holds:
\be 
\label{prop_id}
\ba{l}
\dps
\lim_{\varepsilon\to0}\int_{0}^{\infty} \frac{du}{u^2}\int_{C(\varepsilon)}dz\;G_{\h_i}(\bx,\bx_i)f(\bx) = \pi\int_{0}^{u_i} \frac{du}{u^2}\int_{z_i-i(u-u_i)}^{z_i+i(u-u_i)}dz\;\widetilde{G}_{h_i}(\bx,\bx_i)f(\bx)
\vspace{2.5mm}  
\\
\dps
\hspace{32mm}+\lim_{\varepsilon\to0}\int_{z_i-iu_i}^{z_i+iu_i}dw_i\int_{0}^{\infty} \frac{du}{u^2}\int_{C(\varepsilon)+C_i(\varepsilon)}dz\;\widehat{G}_{h_i}(\bx,\bx_i,w_i)f(\bx)\,,
\ea
\ee 
where $G_{h_i}(\bx,\bx_i)$, $\widehat{G}_{h_i}(\bx,\bx_i,w_i)$, $\widetilde{G}_{h_i}(\bx,\bx_i)$ are given by \eqref{lem_0_G}, \eqref{lem_1_G}, \eqref{lem_2_G}, respectively.
\end{lemma}
The proof is given in Appendix \bref{app:lemma4}. Note that the integration domains in \eqref{prop_id}  and   \eqref{intro_propid} are different. For the sake of simplicity, the latter identity  is given  for the Poincare patch, $\bx\in\mathbb{R}_{\geq0}\times\mathbb{R}$, see the right-hand side of \eqref{intro_propid}. Indeed, the relation \eqref{intro_propid} is the special case of the relation  \eqref{prop_id} obtained by changing  $\bx_i\to\bx'$, $w_i\to w$, $h_i\to h$ and $C(\varepsilon)+C_i(\varepsilon) \to C$, where $C(\varepsilon)=\RR\symbol{92} U_{\varepsilon}(z')$, see fig. \bref{fig:poch}.  

Furthermore, the general form of the superposition identity \eqref{prop_id} allows one to expand a product of $n$ bulk-to-bulk propagators integrated over a  bulk point using the following lemma, which is a direct generalization of Lemma \bref{lem:PROPID}.
\begin{lemma}[the multipoint superposition identity]
\label{lem:propid_n}
Let $f(\bx)$ be a test function satisfying the conditions of Lemma \bref{lem:PROPTILDE}. Applying Lemma \bref{lem:PROPID} to the product of $n$ bulk-to-bulk propagators $n$ times yields  
\be 
\label{propid_n}
\ba{l}
\dps\int_{0}^{\infty} \frac{du}{u^2} \int_{\RR}dz\;\prod_{i=1}^n G_{\h_i}(\bx,\bx_i) f(\bx) 
\vspace{2.5mm}  
\\
\dps
=\pi \sum_{k=1}^n \int_{0}^{u_k} \frac{du}{u^2}\int_{z_k-i(u-u_k)}^{z_k+i(u-u_k)}dz\; \bigg(\prod_{\substack{i=1\\i\neq k}}^{n} \int_{z_i-iu_i}^{z_i+iu_i}dw_i \, \widehat{G}_{h_i}(\bx,\bx_i)\bigg)\widetilde{G}_{h_k}(\bx,\bx_k) f(\bx) 
\vspace{3mm}  
\\
\dps
+\lim_{\substack{\varepsilon_i\to 0 \\ i = 1,...,n} }\bigg(\prod_{i=1}^n\int_{z_i-iu_i}^{z_i+iu_i}dw_i\bigg) \int_{0}^{\infty}\frac{du}{u^2} \int_{C(\varepsilon_1, ... , \varepsilon_n)}dz\; \bigg(\prod_{k=1}^n\widehat{G}_{\h_k}(\bx,\bx_k,w_k)\bigg) f(\bx)\,,
\ea
\ee 
where the contour $C(\varepsilon_1, ... , \varepsilon_n)$ is given by 
\be 
\label{C_n}
C(\varepsilon_1, ... , \varepsilon_n) = \RR \,\symbol{92}\bigcup_{i=1}^n U_{\varepsilon_i}(z_i)+\sum_{i=1}^nC_i(\varepsilon_i)\;,
\ee
where $U_{\varepsilon_i}(z_i)$ is the $\varepsilon_i$-neighbourhood of the point $z_i$ (see fig. \bref{fig:Cmanybumps}).
\end{lemma}
The proof is given in Appendix \bref{app:coroll2}.  The only significant change here compared to Lemma \bref{lem:PROPID} is the form of the $n$-point contour.  Note that choosing  $f(\bx) = 1$ immediately gives the superposition identity for the $n$-point contact AdS Feynman diagram, the schematic form of which is given in \eqref{half_expansion_n}. 
\begin{figure}
\centering
\includegraphics[scale=1]{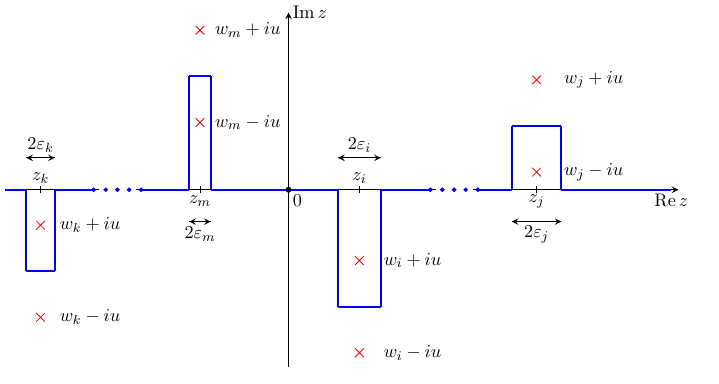}
\caption{The contour $C(\varepsilon_1, ... , \varepsilon_n)$ on the complex $z$-plane. }
\label{fig:Cmanybumps}
\end{figure}

\begin{figure}
\centering
\includegraphics[scale=0.6]{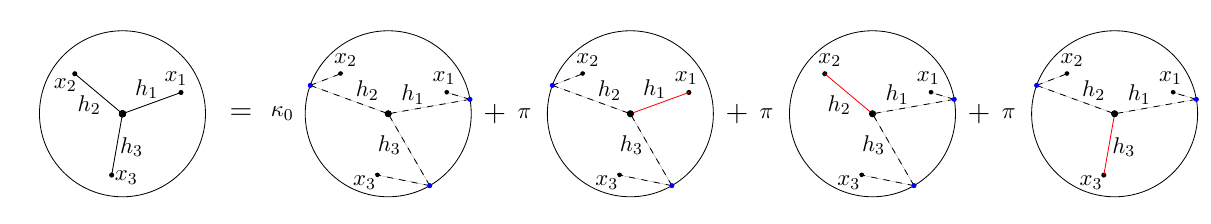}
\caption{The decomposition of the 3-point AdS Feynman diagram into diagrams with modified propagators \eqref{GGG_modified}. The black lines are the bulk-to-bulk propagators $G_h(\bx,\bx')$, the red lines are the modified propagators $\widetilde{G}_h(\bx,\bx')$, the dashed broken  lines are the modified  propagators $\widehat{G}_h(\bx,\bx',w)$. The integrations over $w$ are shown by cusps  located on the boundary.}
\label{fig:relation}
\end{figure}

The (multipoint) superposition identity allows one to decompose  AdS Feynman diagrams into  diagrams with modified propagators. In the case of the $3$-point AdS Feynman diagram this decomposition  reads
\be 
\label{GGG_modified}
\ba{l}
\dps
\int_{\text{AdS}_2}d^2\bx\sqrt{g(\bx)}\, G_{\h_1}(\bx,\bx_1) G_{\h_2}(\bx,\bx_2)G_{\h_3}(\bx,\bx_3)
\vspace{3mm}
\\
\dps
=\coef_0  \prod_{k = 1}^3 \int_{z_k-iu_k}^{z_k+iu_k}dw_k \;\int_0^{\infty} \frac{du}{u^2} \int_Cdz\; \prod_{i=1}^3 \widehat{G}_{h_i}(\bx,\bx_i,w_i)
\vspace{3mm}
\\
\dps
+\pi \bigg(\int_0^{u_1}\frac{du}{u^2}\int_{z_1-i(u-u_1)}^{z_1+i(u-u_1)}dz\; \widetilde{G}_{\h_1}(\bx,\bx_1) \prod_{i=2}^{3} \int_{z_i-iu_i}^{z_i+iu_i}dw_i \, \widehat{G}_{h_i}(\bx,\bx_i)  + (1 \leftrightarrow 2)+ (1 \leftrightarrow 3)\bigg),
\ea
\ee
which directly follows from \eqref{propid_n}. This is the detailed  form of the relation \eqref{half_expansion}. The properties of each term on the right-hand side of the relation were discussed in section \bref{sec:new}. The graphical representation of \eqref{GGG_modified}  in terms of modified bulk-to-bulk propagators is shown in  fig. \bref{fig:relation}. 

Note that the (multipoint) superposition identity does not produce infinite sums of double-trace terms, they are  further generated by applying  the splitting identity \eqref{G_tilde_as_vertex}. Following the algorithm for decomposing  the $3$-point AdS Feynman diagram discussed in section \bref{sec:new}, one obtains the known decomposition of the $3$-point AdS Feynman diagram into the AdS vertex functions \eqref{GGG_rel}:   
\be
\label{GGG_modified2}
\ba{l}
\dps
\text{right-hand side  of}\; \eqref{GGG_modified} =\coef_0  \prod_{k = 1}^3 \int_{z_k-iu_k}^{z_k+iu_k}dw_k\;\int_0^{\infty}\frac{du}{u^2}\int_Cdz\;\prod_{i=1}^3\widehat{G}_{h_i}(\bx,\bx_i,w_i)
\vspace{3mm}
\\
\dps
+ \frac{1}{8\pi^2}\bigg(\frac{(-)^{\h_2+\h_3}}{\sin(2\pi(\h_2+\h_3))} \sum_{n=0}^{\infty} \frac{a(\h_1;\h_2,\h_3;n)}{\alpha'_{\h_2\h_3,n}}\,  \int_{z_k-iu_k}^{z_k+iu_k}dw_k\; \oint_{0} \frac{du}{u^2} \oint_{P[w_1-iu,w_1+iu]} dz\, 
\vspace{3mm}
\\
\dps
\times\,\widehat{G}_{\dth_{23|n}}(\bx,\bx_1,w_1) \widehat{G}_{\h_2}(\bx,\bx_2,w_2) \widehat{G}_{\h_3}(\bx,\bx_3,w_3) + (1 \leftrightarrow 2)+ (1 \leftrightarrow 3)\bigg)
\vspace{3mm}
\\
\dps
= \coef_0\,\cV_{\h_1\h_2\h_3}(\bx_1,\bx_2,\bx_3)
+\bigg(\sum_{n=0}^{\infty}\coef^{^{(1|23)}}_n\cV_{\dth_{23|n} \h_2\h_3}(\bx_1,\bx_2,\bx_3) + (1 \leftrightarrow 2)+ (1 \leftrightarrow 3)\bigg),
\ea
\ee 
where the contours $C$ and $P$ are shown in fig. \bref{fig:contour_KKK} and fig. \bref{fig:poch} ({\bf b}). The first equality is obtained by applying  the splitting identity \eqref{G_tilde_as_vertex} to the terms in the third line of \eqref{GGG_modified}. The last equality is obtained by observing that the first line in \eqref{GGG_modified2} is just the integral representation of the AdS vertex function \eqref{3ptVertexInterm}; similarly, the second  and third lines  are given by \eqref{coroll_vertex}.

\section{Conclusion and outlooks}
\label{sec:conclusion}

In this work, we have shown that AdS Feynman diagrams can be decomposed into Wilson network matrix elements, which serve as a kind of basis functions. We have considered the simplest case of the scalar 3-point AdS Feynman diagram in Euclidean AdS$_2$ space, demonstrating that  such a decomposition involves both a single-trace term  and infinite sums of double-trace terms (terminology derived from  the standard running conformal dimensions), thereby mimicking the conformal block decomposition of Witten diagrams. There are two key ingredients in this construction: first, we utilize  AdS integral propagator identities that transform  the AdS Feynman diagram into an  infinite sum of terms; second,  each term can be represented as a particular Wilson network matrix element.    

Our results can be extended in several directions. First, one may generalize the proposed method  to the $n$-point case with intermediate channels. The first interesting case here is the 4-point exchange AdS Feynman diagram. It is now evident that similar decompositions into Wilson networks can be achieved by applying the superposition identity to the internal lines and finding the $n$-point splitting identity that would relate the $n$-point modified AdS Feynman diagrams and $n$-point AdS vertex functions.  It is also interesting to see how the Wilson  network decomposition can be extended to arbitrary spin fields in $d$ dimensions. Specifically, the $d$-dimensional Wilson line networks and their near-boundary asymptotics  were  considered in \cite{Bhatta:2018gjb} (for more on higher-dimensional spin-networks in the present context see \cite{Penrose} and e.g.  \cite{Ammon:2025cdz}). Of course, the search for various AdS propagator identities in both Euclidean and Lorentzian AdS$_{d+1}$ space is interesting in itself.

\vspace{3mm} 

\noindent \textbf{Acknowledgements.} We are  grateful to  Semyon Mandrygin for useful discussions. Our work was partially supported by the Foundation for the Advancement of Theoretical Physics and Mathematics “BASIS”.

\appendix

\section{Analytic continuation of the 3-point Witten diagram}
\label{app:Witten_analytic}

Let us  consider the $3$-point Witten diagram \cite{Freedman:1998tz} 
\be 
\label{analyt_start}
\frac{1}{\coef_0}\int_{0}^{\infty}\frac{du}{u^2}\int_{\RR}dz\; K_{\h_1}(\bx,w_1)K_{\h_2}(\bx,w_2)K_{\h_3}(\bx,w_3) = \big\langle{\cO_{\h_1}(w_1)\cO_{\h_2}(w_2)\cO_{\h_3}(w_3)}\big\rangle\,,
\ee
where $\bx=(u,z)$, $w_i\in\RR$, $w_i\neq w_j$, $\coef_0$ is defined in \eqref{d_coef},   $h_i > \half$ satisfy  the triangle inequalities \eqref{triangle_identity}. The integral converges only when $w_i$ are real, whereas the HKLL procedure requires integrating \eqref{analyt_start} over complex $w_i$ with a particular kernel (see section \bref{sec:single}). To this end,  
one first considers the following integral
\be 
\label{eps_int}
\ba{c}
\dps
I_{\RR}(w_1,w_2,w_3;\varepsilon)  \equiv \int_{\varepsilon}^{\infty}\frac{du}{u^2}\int_{\RR}dz\; K_{\h_1}(\bx,w_1)K_{\h_2}(\bx,w_2)K_{\h_3}(\bx,w_3)
\vspace{2mm}
\\
\dps
= \int_{\varepsilon}^{\infty}\frac{du}{u^2}u^{\h_1+\h_2+\h_3}\int_{\mathbb{R}}dz\prod_{i=1}^3(u^2 + (w_i-z)^2)^{-h_i}\,,
\ea
\ee
where $\varepsilon > 0$, $w_i\in\CC$, $\re(w_i)\neq\re(w_j)$ for any $i\neq j$, $h_i$ are subject to the same restrictions as in \eqref{analyt_start}, and we substituted the bulk-to-boundary propagator \eqref{bulk-to-boundary}. As the original integral \eqref{analyt_start} is obtained from $I_{\RR}(w_1,w_2,w_3;\varepsilon)$ by taking the limit $\varepsilon\to 0$, the analytic continuation of \eqref{eps_int} also provides the analytic continuation of the original integral by taking the limit $\varepsilon\to 0$.  

The integral $I_{\RR}(w_1,w_2,w_3;\varepsilon)$ converges only when $|\im(w_i)|< \varepsilon$. This follows from that the integrals in \eqref{eps_int} converge iff singularities of the integrand do not lie in the integration domain of $I_{\RR}$. If $|\im(w_i)|\geq \varepsilon$ for some $i$, then a singularity  of the integrand located at $(u,z) = (|\im(w_i)|,\re(w_i))$ lies in the integration domain $(u,z)\in[\varepsilon,\infty)\times \RR$, thereby producing the divergence. 

\begin{figure}
\centering
\includegraphics[scale=0.8]{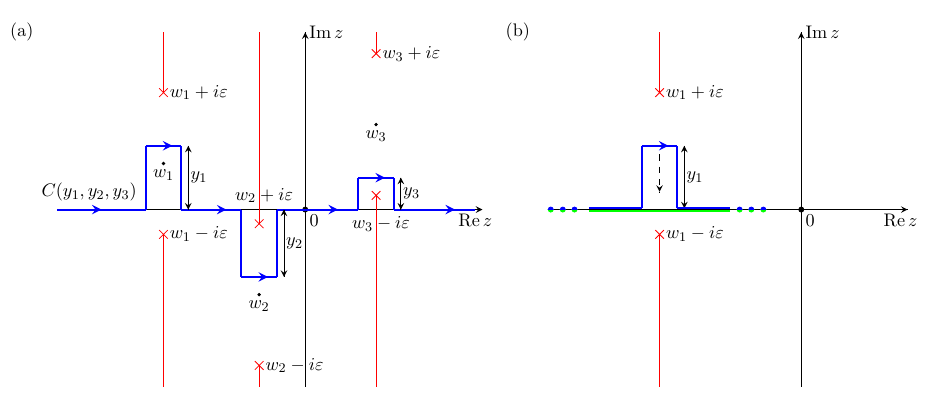}
\caption{{\bf (a)} The contour $C(y_1,y_2,y_3)$  on the complex $z$-plane $\mathbb{C}$. The red crosses and lines show possible singularities of the integrand in \eqref{eps_int}, i.e. if $z$ is taken on the red line, then  $\exists u\in[\varepsilon,\infty)$ such that the integrand is singular in $(u,z)$. Note that  for such a choice of  $w_i$ and $\varepsilon$ the integral $I_{\RR}$ diverges because  possible singularities cross the integration contour which is the real line $z\in\RR$ in this case. Changing the line contour in $I_{\RR}$ to the contour $C(y_1,y_2,y_3)$ allows one to avoid crossing singularities and, therefore, analytically continue $I_{\RR}$ to any complex $w_i$. {\bf (b)} The blue line corresponds to the contour $C(y_1,y_2,y_3)$, the green line corresponds to the line contour. The dashed arrow describes a deformation of $C(y_1,y_2,y_3)$ to the line contour. Note that the restriction $y_1 - \varepsilon<\im(w_i) < \varepsilon$ ensures that the possible singularities do not lie on any contour.}
\label{fig:branch_cuts}
\end{figure}

In order to analytically continue  $I_{\RR}(w_1,w_2,w_3;\varepsilon)$ to the domain $|\im(w_i)|\geq \varepsilon$ one considers the following integral
\be 
\label{eps_mod}
I_{C(y_1,y_2,y_3)}(w_1,w_2,w_3;\varepsilon)  \equiv \int_{\varepsilon}^{\infty}\frac{du}{u^2}\int_{C(y_1,y_2,y_3)}dz\; K_{\h_1}(\bx,w_1)K_{\h_2}(\bx,w_2)K_{\h_3}(\bx,w_3)\,,
\ee
where $y_i\in\RR$ and the contour $C(y_1,y_2,y_3)$ is shown in fig. \bref{fig:branch_cuts}. Then,  the singularities of the integrand in \eqref{eps_mod} do not lie in the integration domain only when $|\im(w_i)-y_i|< \varepsilon$. Thus, \eqref{eps_mod} provides the analytic continuation of $I_{\RR}$ in $w_i\in\CC$, such that $|\im(w_i)-y_i|< \varepsilon$, i.e. by choosing  different parameters $y_i$ in the integration contour  one obtains the analytic continuation for different domains of $w_i$. There are two distinct cases in the proof of this statement: (1) $|y_i|< 2\varepsilon$; (2) $|y_i|\geq2 \varepsilon$. To simplify  the proof we focus  on the analytic continuation in one variable $w_i$ only, e.g. $i=1$ and $y_2=y_3=0$; the proof for other variables is similar.

In the first case $|y_1|< 2\varepsilon$, without  loss of generality, we fix $y_1 > 0$. The case $y_1 < 0$ can be proved similarly,  and the case $y_1=0$ is trivial since $C(0,0,0) = \RR$ and $I_{C(0,0,0)} = I_{\RR}$. To see that \eqref{eps_mod} provides analytic continuation of  \eqref{eps_int}, one shows that the  integrals coincide in the domain where they both converge. The integral $I_{C}$ converges when $y_1 - \varepsilon < \im(w_i) < y_1+\varepsilon$, whereas  the integral $I_{\RR}$ converges when $- \varepsilon < \im(w_i) < \varepsilon$. Due to the restriction $0<y_1< 2\varepsilon$ these domains of convergence have non-empty overlap $y_1 - \varepsilon<\im(w_i) < \varepsilon$. $I_{\RR}$ and $I_{C}$  coincide in this domain, as there exists a continuous deformation that transforms the contour $C(y_1,y_2,y_3)$ to the line contour without crossing singularities or branch cuts of the integrand, see fig. \bref{fig:branch_cuts} {\bf (b)}. This proves the analytic
continuation for $|y_1|< 2\varepsilon$.

In the second case $y_1\geq2 \varepsilon$, where we also fixed $y_1 > 0$, the overlap of domains of convergence of $I_{C}$ and $I_{\RR}$ is empty. To show this, one introduces a chain of analytically continued integrals starting from the original integral $I_\RR$ and ending  with  the integral $I_{C(y_1,y_2,y_3)}$, where a non-empty overlap of domains of convergence exists only for nearest neighbours. This can be done by introducing $n$ auxiliary variables $\alpha_k = ky_1/n$ such that $\alpha_k - \alpha_{k-1} < 2\varepsilon$ and defining $n$ integrals $I_{C(\alpha_k,y_2,y_3)}$. To show that  $I_{C(\alpha_k,y_2,y_3)}$  analytically  continues  $I_{C(\alpha_{k-1},y_2,y_3)}$ one applies the same arguments as in the case $|y_1|< 2\varepsilon$, because  the restriction $\alpha_k - \alpha_{k-1} < 2\varepsilon$ ensures that the domains of convergence of these integrals intersect. The contour $C(\alpha_k,y_2,y_3)$ in this domain can be continuously deformed  to the contour $C(\alpha_{k-1},y_2,y_3)$, this can be done as in fig. \bref{fig:branch_cuts} {\bf (b)}. Thus, $I_{C(\alpha_k,y_2,y_3)} = I_{C(\alpha_{k-1},y_2,y_3)}$ in the domain where both integrals converge. This proves the analytic continuation for $|y_1|\geq 2\varepsilon$.

We have shown  that the analytic continuation of  \eqref{eps_int} to any $w_1\in \CC$ can be achieved by changing the integration contour. Repeating the proof for other  variables $w_2$ and $w_3$, one obtains the following analytic continuation of the integral \eqref{eps_int}:
\be 
\label{z_analyt_integral}
\int_{\varepsilon}^{\infty}\frac{du}{u^2}\int_{C}dz\; K_{\h_1}(\bx,w_1)K_{\h_2}(\bx,w_2)K_{\h_3}(\bx,w_3)\,,
\ee
where $w_i \neq w_j$ for any $i\neq j$ and the contour $C \equiv C(\im(w_1),\im(w_2),\im(w_3))$ is shown in  fig. \bref{fig:contour_KKK}. Note that the variables $y_i$ in  $C(y_1,y_2,y_3)$ are chosen to be equal to $\im(w_i)$. In this way, one ensures that the domain of convergence of \eqref{z_analyt_integral} does not depend on $\varepsilon$: it is given by $|\im(w_i) - y_i|< \varepsilon$, and, therefore, in the case $y_i = \im(w_i)$, the restriction is satisfied for any $\varepsilon>0$ and $w_i\in\CC$. By taking the limit $\varepsilon\to 0$ in \eqref{z_analyt_integral} one obtains the analytically continued $3$-point Witten diagram
\be 
\label{analyt_end}
\frac1\coef_0\, \int_{0}^{\infty} \frac{du}{u^2}\int_{C}dz\;K_{\h_1}(\bx,w_1)K_{\h_2}(\bx,w_2)K_{\h_3}(\bx,w_3) = \big\langle{\cO_{\h_1}(w_1)\cO_{\h_2}(w_2)\cO_{\h_3}(w_3)}\big\rangle\,,
\ee
where $w_i\in \CC$ and $\re(w_i)\neq\re(w_j)$ for any $i\neq j$.

\paragraph{Special weights.} In the case when $h_1 = h_2+h_3+2n$, $n\in \mathbb{N}_0$ and  $h_2,h_3 \in \mathbb{N}$, the $3$-point Witten diagram \eqref{KKK} diverges, i.e. it cannot  represent the $3$-point conformal correlation function. In this case, one uses the following integral representation of the $3$-point conformal correlation function that can be checked by calculating both contour integrals on the right-hand side:
\be 
\label{spec_h}
\ba{l}
\dps
\big\langle{\cO_{\h_1}(w_1)\cO_{\h_2}(w_2)\cO_{\h_3}(w_3)}\big\rangle
\vspace{2.5mm}  
\\
\dps
=  \frac{\pref_{\h_1 \h_2\h_3}}{\alpha'_{\h_2\h_3,n}}\frac{1}{(2\pi i)^2}\oint_{u=0} \frac{du}{u^2} \oint_{z=w_1+iu} dz\, K_{\h_1}(\bx,w_1)K_{\h_2}(\bx,w_2)K_{\h_3}(\bx,w_3)\,,
\ea 
\ee 
where $\bx = (u,z)$, the $\alpha'$-coefficient is given by \eqref{alpha_prime}, $\pref_{\h_1 \h_2\h_3}$ is the structure constant \eqref{prefactor}, and  $w_i\in \CC$,  $\re(w_i)\neq\re(w_j)$ for any $i\neq j$. 

In the case when $h_1 = h_2+h_3+2n$, $n\in \mathbb{N}_0$ and  $h_2,h_3 \in \RR$, the formula \eqref{spec_h} can be analytically continued as follows 
\be 
\label{3pt_poch}
\ba{l}
\dps
\big\langle{\cO_{\h_1}(w_1)\cO_{\h_2}(w_2)\cO_{\h_3}(w_3)}\big\rangle
\vspace{2.5mm}  
\\
\dps
= \frac{\pref_{\h_1 \h_2\h_3}}{\alpha'_{\h_2\h_3,n}}\frac{1}{8\pi^2 i}\frac{(-)^{\h_2+\h_3}}{\sin(2\pi\h_1)}\oint_{u=0} \frac{du}{u^2} \oint_{P[w_1+iu,w_1-iu]} dz\;K_{\h_1}(\bx,w_1)K_{\h_2}(\bx,w_2)K_{\h_3}(\bx,w_3)\,,
\ea 
\ee 
where $P[w_1+iu,w_1-iu]$ is the Pochhammer contour around the points $w_1\pm iu$, see fig. \bref{fig:poch}.

\section{Proofs of lemmas}
\label{app:lemmas}

\subsection{Lemma \bref{lem:PROPSIMPLE}}
\label{app:lemma1}

To prove the conversion identity we explicitly calculate the integral in the right-hand side of \eqref{KK_as_bb}, which we call the reconstructed propagator,
\be 
\label{G_rec}
\int_{z'-iu'}^{z'+iu'}dw\;\widehat{G}_{h}(\bx,\bx',w) \equiv G_h^{\text{rec}}(\bx,\bx')\,,
\ee
where $\widehat{G}_{h}(\bx,\bx',w)$ is given in \eqref{propagators}, and then show that it coincides with the bulk-to-bulk propagator. Substituting $\widehat{G}_{h}(\bx,\bx',w)$ into \eqref{G_rec} one obtains 
\be 
G^{\text{rec}}_h(\bx,\bx')  = \frac{-2i}{4^{\h}}\frac{\Gamma(2\h)}{\Gamma(\h)\Gamma(\h)}\int_{z'-iu'}^{z'+iu'}d w\;u'^{\,1-h}u^{h}(u'^{\,2}+(z'-w)^2)^{h-1}(u^2+(z-w)^2)^{-h}\,.
\ee
Making the change $w=z'+iu'(2t-1)$ one obtains
\be 
\label{lemma1_interm}
\ba{l}
\dps
G_h^{\text{rec}}(\bx,\bx')  = \frac{\Gamma(2\h)}{\Gamma(\h)\Gamma(\h)}\, \left(\frac{uu'}{(u'+u+i(z-z'))(u'-u+i(z-z'))}\right)^{h}
\vspace{2.5mm}  
\\
\dps
\times
\int_0^1dt\;t^{h-1}(1-t)^{h-1}\left(1-t\frac{2u'}{u+u'+i(z-z')}\right)^{-h}\left(1-t\frac{2u'}{u'-u+i(z-z')}\right)^{-h} 
\vspace{2.5mm}  
\\
\dps
=\left(\frac{uu'}{(u+u'+i(z-z'))(u'-u+i(z-z'))}\right)^{h}
\vspace{2.5mm}  
\\
\dps
\times
F_1 \left[\begin{array}{ccc}
    h,& h, & h  \\
     &2 h \end{array};\frac{2u'}{u'+u+i(z-z')},\frac{2u'}{u'-u+i(z-z')}\right],
\ea
\ee
where we used the integral representation of the first Appell function \cite{Bateman:100233}
\be 
\label{def_appell1}
F_\text{1} \left[\begin{array}{cccc}
     a,&b_1, &b_2  \\
     &c \end{array};z_1,z_2\right] =\frac{\Gamma(c)}{\Gamma(a)\Gamma(c-a)}\int_0^1dx\;x^{a-1}(1-x)^{c-a-1}(1-z_1x)^{-b_1}(1-z_2x)^{-b_2}\,.
\ee 
Here, $z_1,z_2\in\CC\symbol{92}[1,\infty)$ that in \eqref{lemma1_interm} is  equivalent to imposing the following restrictions: 
\be
\label{restrict}
\re(z)\neq \re(z') \quad \text{or} \quad  u>u'-|\im(z-z')|, \quad \text{where} \;\; u,u'\in\RR_+ \;\; \text{and} \;\; z,z'\in\CC\,.
\ee 
Using the following identity \cite{Bateman:100233}
\be 
\label{F1_transform}
F_1 \left[\begin{array}{ccc}
     a,&b_1, &b_2  \\
     &c \end{array};z_1,z_2\right] =(1-z_1)^{-a} F_1 \left[\begin{array}{ccc}
     a,&c-b_1-b_2, &b_2  \\
     &c \end{array};\frac{z_1}{z_1-1},\frac{z_1-z_2}{z_1-1}\right],
\ee 
one obtains that the reconstructed propagator is the standard bulk-to-bulk propagator:
\be 
\label{KK_as_bb_prefinal}
\ba{l}
\dps
G_h^{\text{rec}}(\bx,\bx') = \left(\frac{u u'}{(u+u')^2+(z-z')^2}\right)^{h}
F_1 \left[\begin{array}{ccc}
    h,& h, & 0 \\
     &2 h \end{array};\frac{4u u'}{(u+u')^2+(z-z')^2}, \frac{2u'}{u'+u-i(z-z')}\right]   
\vspace{2.5mm}  
\\
\dps
=\left(\frac{\xi(\bx,\bx')}{2}\right)^{\h}{}_2F_1\left(\frac{\h}{2},\frac{\h}{2}+\half; \h+\half\Big|\xi(\bx,\bx')^2\right) = G_{\h}(\bx,\bx')\,,
\ea
\ee
where $\xi(\bx,\bx')$  is defined in \eqref{bulk-to-bulk}. Here, we used that the first Appell function in \eqref{KK_as_bb_prefinal} becomes the Gauss hypergeometric function and then applied the quadratic transformation \cite{Bateman:100233} 
\be 
\label{quadratic}
\ba{l}
\dps
{}_2F_1(a, b; 2a|z)=\left(1-\frac{z}{2}\right)^{-b}{}_2F_1\left(\frac{b}{2},\frac{b}{2}+\half;a+\half\Big|\frac{z^2}{(2-z)^2}\right)\,,
\ea
\ee 
to obtain the bulk-to-bulk propagator \eqref{bulk-to-bulk}. This proves Lemma \bref{lem:PROPSIMPLE}.

\subsection{Lemma \bref{lem:PROPTILDE}}
\label{app:lemma2}

Let us  consider the integral on the left-hand side of \eqref{R_def}:
\be 
\label{R_def_app}
R(\bx_i, \varepsilon) \equiv  \int_{z_i-iu_i}^{z_i+iu_i}\hspace{-2mm}dw\int_{0}^{\infty} \frac{du}{u^2}\int_{C_i(\varepsilon)}dz\,\widehat{G}_{h_i}(\bx,\bx_i,w) f(\bx) \,,
\ee 
where $\varepsilon \to 0$ and  the contour $C_i(\varepsilon)$ is shown in fig. \bref{fig:C_i}. Everywhere below we assume  that a test function $f(\bx)$ is holomorphic  on a domain $U \times V \subset \CC^2$, such that $U\subset \RR $ and $V \subset C_i(\varepsilon)$ for any $\varepsilon \to 0$. We also assume  that the integrand in \eqref{R_def_app} is finite at $u=0$ for $\forall z\in C_i(\varepsilon)$ that restricts $f(\bx)$ as follows: $u^{-h_i}f(\bx) = \cO(u)$ at $u\to 0$.\footnote{In the case when a test function is chosen to be a product of two bulk-to-bulk propagators, $f(\bx) = G_{\h_j}(\bx,x_j)G_{\h_k}(\bx,x_k),$ $ i\neq j\neq k\neq i$, this restriction is equivalent to the triangle inequality $h_j+h_k> h_i$.}

Splitting  the integration contour $C_i(\varepsilon)$  into three line segments one obtains
\be 
\label{integral_interm_pre}
\ba{l}
\dps
R(\bx_i, \varepsilon) = \int_{z_i-iu_i}^{z_i+iu_i}\hspace{-2mm}dw\int_{0}^{\infty} \frac{du}{u^2}\int_{w_i-\varepsilon}^{w_i+\varepsilon}dz\;\widehat{G}_{h_i}(\bx,\bx_i,w) f(\bx) 
\vspace{2.5mm}  
\\
\dps
\hspace{15mm}
+ \int_{z_i-iu_i}^{z_i+iu_i}\hspace{-2mm}dw\int_{0}^{\infty} \frac{du}{u^2}\int_{z_i - \varepsilon}^{w_i- \varepsilon}dz\;\widehat{G}_{h_i}(\bx,\bx_i,w) f(\bx) 
\vspace{2.5mm}  
\\
\dps
\hspace{15mm}
- \int_{z_i-iu_i}^{z_i+iu_i}\hspace{-2mm}dw\int_{0}^{\infty} \frac{du}{u^2}\int_{z_i + \varepsilon}^{w_i+ \varepsilon}dz\;\widehat{G}_{h_i}(\bx,\bx_i,w) f(\bx) \,.
\ea
\ee 
The first term is proportional to $\varepsilon$ because a length of the contour equals $2\varepsilon$ and the integrand is holomorphic in the integration domain. Then, by changing $z$ to $z'=z\pm \varepsilon$ in the second and  third lines one represents \eqref{integral_interm_pre} up to  $O(\varepsilon)$ terms as follows 
\be 
\label{integral_interm}
R(\bx_i, \varepsilon) = 
\int_{z_i-iu_i}^{z_i+iu_i}\hspace{-2mm}dw\int_{0}^{\infty} \frac{du}{u^2}\int_{z_i}^{w_i}dz\;\big(\widehat{G}_{h_i}(z-\varepsilon,u,\bx_i,w) - \widehat{G}_{h_i}(z+\varepsilon,u,\bx_i,w)\big) f(\bx) +O(\varepsilon)\,,
\ee 
where: (1) we took into account that $f(\bx)$ is holomorphic; (2) we  expanded  $f(u,z'\pm\varepsilon)$ into the Taylor series around $\bx = (u,z')$; (3) we omitted terms linear in $\varepsilon$; (4) we renamed $z'\to z$, for convenience. The integrand is a test function times a discontinuity of the modified propagator. 

Before calculating \eqref{integral_interm} explicitly, let us find domains of $z$, $u$, $w$ where  the discontinuity is non-vanishing. Using the modified propagator \eqref{propagators} and the following relation 
\be 
\label{disc_power}
(x+i\varepsilon)^\alpha-(x-i\varepsilon)^\alpha = \theta(-x)(e^{\pi i \alpha}-e^{-\pi i \alpha})(x+i\varepsilon)^\alpha+O(\varepsilon)\,, \quad \alpha\in\mathbb{R}\,,
\ee 
one finds that 
\be 
\label{disc_def}
\ba{c}
\dps
\text{Disc}\big(\widehat{G}_{h_i}(z\mp \varepsilon,u,\bx_i,w)\big) = \widehat{G}_{h_i}(z- \varepsilon,u,\bx_i,w) - \widehat{G}_{h_i}(z+\varepsilon,u,\bx_i,w)
\vspace{2.5mm}  
\\
\dps
=\text{sign}(iz-iw)\, \theta((iz-iw)^2-u^2) (e^{-\pi i h_i}-e^{\pi i h_i}) \, \widehat{G}_{h_i}(z+ \varepsilon,u,\bx_i,w)+O(\varepsilon)\,,
\ea
\ee
where the sign function is defined as $\text{sign}(x) = 2\theta(x)-1$. Here, $iz-iw\in\mathbb{R}$, which follows from the integration domains of $w$ and $z$ in \eqref{integral_interm}, ensuring that arguments of the sign and step functions are real. The resulting restriction on $z$, $u$, $w$ is given by the step function in \eqref{disc_def}, i.e. $(iz-iw)^2-u^2 \geq0$. 

Returning to the integral \eqref{integral_interm}, one substitutes $\widehat{G}_{h_i}(\bx,\bx_i,w)$ \eqref{propagators}, takes into account the restriction $(iz-iw)^2-u^2\geq0$ for $u$ and $z$ (otherwise the integrand is zero), makes the change $z=z_i+iv$, $w=z_i+it$, and then interchanges integrations over $t$ and $v$:
\be 
\label{disc_integral_2}
\ba{l}
\dps
R(\bx_i, \varepsilon) = \frac{2i}{4^{\h_i}}\frac{\Gamma(2\h_i)}{\Gamma(\h_i)\Gamma(\h_i)}u_i^{1-h_i}\int_{0}^{u_i} \frac{du}{u^2}\;u^{h_i}\int_0^{u_i-u}dv\;\big(I_1(u,v,\varepsilon) - I_1(u,v,-\varepsilon)\big)f(z_i+iv,u)
\vspace{2.5mm}  
\\
\dps
- \frac{2i}{4^{\h_i}}\frac{\Gamma(2\h_i)}{\Gamma(\h_i)\Gamma(\h_i)}u_i^{1-h_i}\int_{0}^{u_i} \frac{du}{u^2}\;u^{h_i}\;\int_{u-u_i}^0 dv\;\big(I_2(u,v,\varepsilon) - I_2(u,v,-\varepsilon)\big)f(z_i+iv,u)+O(\varepsilon)\,,
\ea
\ee 
where 
\be 
\label{I_def}
\ba{l}
\dps
I_1(u,v,\varepsilon) = \int_{v}^{u_i}dt\;(u_i^2-t^2)^{h_i-1}(u^2 + (iv-it-\varepsilon)^2)^{-h_i}\,,
\vspace{2.5mm}  
\\
\dps
I_2(u,v,\varepsilon) = \int_{-u_i}^{v}dt\;(u_i^2-t^2)^{h_i-1}(u^2 + (iv-it-\varepsilon)^2)^{-h_i}\,.
\ea 
\ee 
The integrals are split due to the Heaviside step function and the interchange of the order of integration. Note that the domain of $t$ is not restricted as the integrals \eqref{I_def} are easier to calculate without imposing such a restriction. 

Now, we calculate $I_1(u,v,\varepsilon)$ by changing  $t = v+(u_i-v)s$, $s\in[0,1]$, and using the integral representation of the Lauricella function \cite{Matsumoto_2020}
\be 
\label{def_lauricella}
F^n_\text{D} \left[\begin{array}{cccc}
     a,&b_1, \dots, &b_n  \\
     &c \end{array};z_1,\dots,z_n\right] =\frac{\Gamma(c)}{\Gamma(a)\Gamma(c-a)}\int_0^1dx\;x^{a-1}(1-x)^{c-a-1}\prod_{i=1}^n(1-z_ix)^{-b_i}\,.
\ee 
The result is 
\be
\label{interm_integral}
\ba{l}
\dps
I_1(u,v,\varepsilon)=\frac{\Gamma(h_i)}{\Gamma(h_i+1)}\left(\frac{u_i-v}{(u+i\varepsilon)(u-i\varepsilon)}\right)^{h_i}(u_i+v)^{h_i-1}
\vspace{2.5mm}  
\\
\dps
\hspace{20mm}\times\,
F^3_\text{D} \left[\begin{array}{cccc}
     1, &1-h_i,&h_i, &h_i  \\
     & h_i+1 \end{array};\frac{v-u_i}{u_i+v},\frac{u_i-v}{u+i\varepsilon},\frac{v-u_i}{u-i\varepsilon}\right].
\ea
\ee 
The Lauricella function on the right-hand side can be transformed by using the following relation  \cite{Matsumoto_2020}
\be 
\ba{l}
\dps
F^3_\text{D} \left[\begin{array}{cccc}
     a,&b_1, &b_2, &b_3  \\
     & &c \end{array};z_1,z_2,z_3\right]
\vspace{2.5mm}  
\\
\dps
= (1-z_3)^{-a}F^3_\text{D} \left[\begin{array}{cccc}
    a, &b_1,&b_2, &c-\sum_{i=1}^3b_i  \\
     & &c \end{array};\frac{z_1-z_3}{1-z_3},\frac{z_2-z_3}{1-z_3},\frac{z_3}{z_3-1}\right]\,,
\ea
\ee 
where we note that $c$, $b_i$ in \eqref{interm_integral} satisfy $c-\sum_{i=1}^3b_i = 0$, and,   therefore, the Lauricella function reduces to the first Appell function,
\be
\label{I1_res}
\ba{c}
\dps
I_1(u,v,\varepsilon)=
\frac{\Gamma(h_i)}{\Gamma(h_i+1)}\left(\frac{u_i-v}{u+i\varepsilon}\right)^{h_i}\left(\frac{u_i+v}{u-i\varepsilon}\right)^{h_i-1}\left(u+u_i-v-i\varepsilon\right)^{-1}
\vspace{2.5mm}  
\\
\dps
\times F_\text{1    } \left[\begin{array}{ccc}
     1, &1-h_i,&h_i \\
     & h_i+1 \end{array};\frac{u_i-v}{u_i+v}\frac{u_i-u+v+i\varepsilon}{u_i+u-v-i\varepsilon},\frac{2u_i(u_i-v)}{(u+i\varepsilon)(u_i+u-v-i\varepsilon)}\right].
\ea
\ee 

In order to find $I_1(u,v,\varepsilon) - I_1(u,v,-\varepsilon)$ at $\varepsilon\to 0$ one studies the analytical properties of the function $I_1(u,v,\varepsilon)$. The prefactor in \eqref{I1_res} is a continuous function of $\varepsilon$ at $\varepsilon\to 0^{\pm}$ for $u\leq u_i$ and $0\leq v\leq u_i-u$ because the real parts of the arguments of all power functions are greater than zero. The only source of  discontinuity in $I_1(u,v,\varepsilon)$ is the first Appell function $ F_\text{1    } \left[\begin{array}{ccc}
     a, &b_1,&b_2 \\
     & c \end{array};z_1,z_2\right]$, which has two branch cuts $\re(z_1)>1,\, \im(z_1)=0$ and $\re(z_2)>1,\, \im(z_2)=0$. In our case, the restrictions on $u$ and $v$ arising  from the integration domain constrain the arguments of the first Appell function as follows
 \be 
 \ba{l}
 \dps
z_1 = \frac{u_i-v}{u_i+v}\frac{u_i-u+v+i\varepsilon}{u_i+u-v-i\varepsilon}\,,\qquad|z_1| \leq 1 \,;
\vspace{2.5mm}  
\\
\dps
z_2 = \frac{2u_i(u_i-v)}{(u+i\varepsilon)(u_i+u-v-i\varepsilon)}\,,\qquad2>\re z_2\geq 1,\; \im z_2 \sim \varepsilon, \; |1-z_2|<1 \,,
\ea
\ee 
i.e. the discontinuity of $I_1(u,v,\varepsilon)$ is proportional to the discontinuity of the first Appell function at the branch cut $\re(z_2)>1$, $\im(z_2)=0$.

The discontinuity can be found by using the following expansion of the first Appell function at $|z_1|<1$ and $|1-z_2|<1$ \cite{bezrodnykh2017analytic}:
\be 
\label{Appell_exp}
\ba{l}
\dps
 F_\text{1} \left[\begin{array}{ccc}
     a, &b_1,&b_2 \\
     &a+b_2 \end{array}; z_1, z_2\right] = -\frac{\Gamma(a+b_2)}{\Gamma(a)\Gamma(b_2)}\sum_{k=0}^\infty\frac{(a)_k(b_2)_k}{k!k!}(1-z_2)^k\Bigg[\frac{d}{ds}{}_2F_1(b_1,a+s,a, z_1)\bigg|_{s=k}
\vspace{2.5mm}  
\\
\dps
+ \ln(1-z_2) {}_2F_1(b_1,a+k,a, z_1) - h_k^+(b_2,a)\,{}_2F_1(b_1,a+k,a, z_1)\Bigg].
\ea
\ee
Here, $h_k^+(b_2,a)$ is some function of $b_2, a$, and $k$,  whose form is irrelevant for calculating the discontinuity because the  term in \eqref{Appell_exp} containing  $h_k^+(b_2,a)$ is a holomorphic function of $z_1$ and $z_2$ in the domain $|1-z_2|<1$ and $|z_1|<1$. The same applies to  the first line in \eqref{Appell_exp}. The only term contributing to the discontinuity is that one containing the logarithm. Substituting \eqref{Appell_exp} into  $I_1(u,v,\varepsilon) - I_1(u,v,-\varepsilon)$ one obtains 
\be
\ba{l}
\dps
I_1(u,v,\varepsilon) - I_1(u,v,-\varepsilon)= 2i \pi
\left(\frac{u_i-v}{u}\right)^{h_i}\left(\frac{u_i+v}{u}\right)^{h_i-1}\left(u+u_i-v\right)^{-1}
\vspace{2.5mm}  
\\
\dps
\times F_\text{2    } \left[\begin{array}{ccc}
     1, &h_i,&1-h_i \\
     &1, & 1 \end{array};\frac{-u_i+u+v}{u_i+u-v},\frac{u_i-v}{u_i+v}\frac{u_i-u+v}{u_i+u-v}\right]+O(\varepsilon)\,,
\ea
\ee 
where we summed the discontinuity of \eqref{Appell_exp} over  $k$ to obtain the second Appell function, which can be further simplified using   two  identities \cite{Bateman:100233}:
\be 
\ba{l}
\dps
 F_\text{1} \left[\begin{array}{ccc}
     a, &b_1,&b_2 \\
     &a, & a \end{array}; z_1, z_2\right] = (1-z_1)^{-b_1}(1-z_2)^{-b_2}\,{}_2F_{1}\left(b_1,b_2;a\Big|\frac{z_1z_2}{(1-z_1)(1-z_2)}\right),
\vspace{2.5mm}  
\\
\dps
{}_2F_{1}\left(a,1-a;c\Big|z\right) = (1-z)^{c-1}(1-2z)^{a-c}\,{}_2F_{1}\left(\frac{c-a}{2},\frac{c-a+1}{2};c\Big|\frac{4z(1-z)}{(1-2z)^2}\right).
\ea
\ee 
The resulting expression  is given by 
\be
\label{I1_disc}
\ba{l}
\dps
I_1(u,v,\varepsilon) - I_1(u,v,-\varepsilon)
\vspace{2.5mm}  
\\
\dps
=i \pi
u^{-h_i}u_i^{h_i-1}\xi({\bf y},\bx_i)^{1-h_i}\,{}_2F_{1}\left(\half-\frac{h_i}{2},1-\frac{h_i}{2};1\Big|1-\xi({\bf y},\bx_i)^2\right)+O(\varepsilon)\,,
\ea
\ee 
where ${\bf y} = (u,z_i+iv)$ and $\xi({\bf y},\bx_i)$ is defined in \eqref{bulk-to-bulk}. Similarly, one finds the discontinuity of $I_2(u,v,\varepsilon)$:
\be
\label{I2_disc}
\ba{l}
\dps
I_2(u,v,\varepsilon) - I_2(u,v,-\varepsilon)
\vspace{2.5mm}  
\\
\dps
=-i \pi u^{-h_i}u_i^{h_i-1}\xi({\bf y},\bx_i)^{1-h_i}\,{}_2F_{1}\left(\half-\frac{h_i}{2},1-\frac{h_i}{2};1\Big|1-\xi({\bf y},\bx_i)^2\right)+O(\varepsilon)\,.
\ea
\ee 
By substituting \eqref{I1_disc} and  \eqref{I2_disc} into \eqref{disc_integral_2} one obtains the final relation 
\be 
\label{modified_G}
\ba{c}
\dps 
\lim_{\varepsilon\to0}R(\bx_i, \varepsilon) = \int_{z_i-iu_i}^{z_i+iu_i}\hspace{-2mm}dw\int_{0}^{\infty} \frac{du}{u^2}\int_{C_i(\varepsilon)}dz\,\widehat{G}_{h_i}(\bx,\bx_i,w) f(\bx)
\vspace{2.5mm}  
\\
\dps
\hspace{12mm} = \pi \int_{0}^{u_i} \frac{du}{u^2}\int_{z_i-i(u_i-u)}^{z_i+i(u_i-u)}dz\;\widetilde{G}_{h_i}(\bx,\bx_i)f(\bx) \,,
\ea
\ee
where we used the definition of $R(\bx_i, \varepsilon)$ \eqref{R_def_app}, made the change $v = iz_i - iz$, and used  the modified propagator \eqref{propagators}. This proves Lemma \bref{lem:PROPTILDE}. 

\subsection{Lemma \bref{lem:DOUBLE-TRACE}}
\label{app:lemma3}

Let us consider the integral on the left-hand side of \eqref{G_tilde_as_vertex} with two points taken on the boundary $\bx_1=(0,z_1)$, $\bx_2 = (0,z_2)$, which is  multiplied by the factor $\frac{4^{\h_3}}{-2i}\frac{\Gamma(\h_3)\Gamma(\h_3)}{\Gamma(2\h_3)}$:
\be 
\label{G_tilde_app}
\ba{l}
\dps
V(z_1,z_2,\bx_3) \equiv \frac{4^{\h_3}}{-2i}\frac{\Gamma(\h_3)\Gamma(\h_3)}{\Gamma(2\h_3)}\frac{1}{8i\pi^3}\frac{(-)^{\h_1+\h_2}}{\sin(2\pi(\h_1+\h_2))}\sum_{n=0}^{\infty}\frac{a(\h_3;\h_1,\h_2;n)}{\alpha'_{\h_1\h_2,n}}\,
\vspace{2.5mm}  
\\
\dps
\times  
\int_{z_3-iu_3}^{z_3+iu_3} \hspace{-2mm} dw\;\oint_{0} \frac{du}{u^2} \oint_{P[w-iu,w+iu]} \hspace{-2mm} dz\;K_{\h_1}(\bx,z_1)K_{\h_2}(\bx,z_2) \widehat{G}_{\dth_{12|n}}(\bx,\bx_3,w)\,,
\ea
\ee 
where the coefficients $a(h_3;h_1,h_2;n)$, $\alpha'_{\h_1\h_2,n}$ are defined in \eqref{a_coef}, \eqref{alpha_prime}, $P[w-iu,w+iu]$ is the Pochhammer contour, see fig. \bref{fig:poch}, $K_{\h_i}(\bx,z_i)$ is the bulk-to-boundary propagator \eqref{bulk-to-boundary} and $\widehat{G}_{\dth_{12|n}}(\bx,\bx_3,w)$ is the modified propagator \eqref{propagators}. In order to prove Lemma \bref{lem:DOUBLE-TRACE} one shows that \eqref{G_tilde_app} coincides (up to the additional factor of $\frac{4^{\h_3}}{-2i}\frac{\Gamma(\h_3)\Gamma(\h_3)}{\Gamma(2\h_3)}$) with the left-hand side of \eqref{G_tilde_as_vertex}, where two points are taken on the boundary. To this end, we perform a series of  transformations including integrations and infinite summations. 

Substituting $a(h_3;h_1,h_2;n)$, $\alpha'_{\h_1\h_2,n}$, $K_{\h_i}(\bx,z_i)$, $\widehat{G}_{\dth_{12|n}}(\bx,\bx_3,w)$ and changing $z = w-iu+ 2ius $, $w = z_3-iu_3+2iu_3t$ in \eqref{G_tilde_as_vertex} yields
\be 
\label{KGK_tilde_start}
\ba{l}
\dps
V(z_1,z_2,\bx_3) = -\frac{1}{4i\pi^2}\frac{(-)^{\h_3}}{\sin(2\pi(\dth_{12|n}))}
\sum_{n=0}^{\infty}\frac{(\dth_{12|n}-1)}{(\dth_{12|n}-\h_3)(\dth_{12|n}+\h_3-1)} \oint_{0} \frac{du}{u} \int_{0}^{1}dt\;\oint_{P[0,1]} ds \;
\vspace{2.5mm}  
\\
\dps
\times  
\big(u^2-(u+u_3+iz_{13}-2us-2u_3t)^2\big)^{-\h_1}\big(u^2-(u+u_3+iz_{23}-2us-2u_3t)^2\big)^{-\h_2}
\vspace{2.5mm}  
\\
\dps
\times
\big(t(1-t)\big)^{\dth_{12|n}-1}\big(s(1-s)\big)^{-\dth_{12|n}}\left(\frac{u_3}{u}\right)^{2n}.
\ea
\ee 
To integrate over $s$ one uses the analytically continued integral representation of the Lauricella function $F^4_\text{D}$, which is the regular integral representation \eqref{def_lauricella}, where the integration contour is changed to the Pochhammer contour $P[0,1]$. By expanding the resulting  function $F^4_\text{D}$ into a series one obtains
\be 
\ba{l}
\dps
V(z_1,z_2,\bx_3)= \frac{(-)^{h_3}}{\pi}\sum_{n=0}^{\infty}\frac{\Gamma(2\dth_{12|n})}{\Gamma(\dth_{12|n})^2}\frac{(-)^{2\dth_{12|n}}}{(\dth_{12|n}-\h_3)(\dth_{12|n}+\h_3-1)}
\vspace{2.5mm}  
\\
\dps
\times \oint_{0} \frac{du}{u} \int_{0}^{1}dt\; \sum_{k_1,...,k_4=0}^{\infty} \frac{(1-\dth_{12|n})_{K}(h_1)_{k_1}(h_1)_{k_2}(h_2)_{k_3}(h_2)_{k_4}}{(2-2\dth_{12|n})_{K}k_1!k_2!k_3!k_4!}
\vspace{2.5mm}  
\\
\dps
\times 
\big(t(1-t)\big)^{\dth_{12|n}-1}u_3^{2n}u^{K-2n}2^{K} \big(2u+u_3+iz_{13}-2u_3t\big)^{-\h_1-k_1}
\vspace{2.5mm}  
\\
\dps
\times
\big(u_3+iz_{13}-2u_3t\big)^{-\h_1-k_2}\big(2u+u_3+iz_{23}-2u_3t\big)^{-\h_2-k_3}\big(u_3+iz_{23}-2u_3t\big)^{-\h_2-k_4}\,,
\ea
\ee 
where, for convenience,  we denoted $K = k_1+k_2+k_3+k_4$, and used the series representation of  the Lauricella function, 
\be 
\label{series_Lauricella}
F^n_\text{D} \left[\begin{array}{cccc}
     a,&b_1, \dots, &b_n  \\
     &c \end{array};z_1,\dots,z_n\right]  = \sum_{m_1,\dots,m_n =1}^{\infty}\frac{(a)_{\sum_{i=1}^nm_i}}{(c)_{\sum_{i=1}^nm_i}}\prod_{i=1}^n(b_i)_{m_i}\frac{z_i^{m_i}}{m_i!}\,,\quad |z_i| < 1\,.
\ee 
To integrate over $t$ one again uses the integral representation of the Lauricella function $F^4_\text{D}$ and expands it into a series,
\be 
\ba{l}
\dps
\hspace{-12mm}V(z_1,z_2,\bx_3) = \frac{(-)^{h_3}}{\pi}\sum_{n=0}^{\infty}\frac{(-)^{2\dth_{12|n}}}{(\dth_{12|n}-\h_3)(\dth_{12|n}+\h_3-1)} \oint_{0} \frac{du}{u}\;\sum_{\substack{k_1,...,k_4=0\\l_1,...,l_4=0}}^{\infty}\frac{(\dth_{12|n})_{L}}{(2\dth_{12|n})_{L}}
\vspace{2.5mm}  
\\
\dps
\hspace{-12mm}\times
\frac{(1-\dth_{12|n})_{K}}{(2-2\dth_{12|n})_{K}}\frac{(h_1)_{k_1+l_1}(h_1)_{k_2+l_2}(h_2)_{k_3+l_3}(h_2)_{k_4+l_4}}{k_1!k_2!k_3!k_4!l_1!l_2!l_3!l_4!}2^{K+L}u^{K-2n}u_3^{L+2n}
\ea
\ee 
$$
\dps
\times
(2u+u_3+iz_{13})^{-\h_1-k_1-l_1}(u_3+iz_{13})^{-\h_1-k_2-l_2}
(2u+u_3+iz_{23})^{-\h_2-k_3-l_3}(u_3+iz_{23})^{-\h_2-k_4-l_4},
$$
where we denoted $L = l_1+l_2+l_3+l_4$. Note that the convergence of the infinite sums here is guaranteed by the following restrictions on $z_{13}$ and $z_{23}$: $|z_{13}|>2u_3$ and $|z_{23}|>2u_3$. 

The next step is to sum over $n$. By imposing the restriction $|u|> u_3$ on the integration contour in the integral over $u$, which can be achieved by  continuous contour deformation, we ensure that after the change of order of the summation over $n$ and the integration over $u$ the infinite sum over $n$ converges. This sum is in fact the hypergeometric function $_{10}F_{9}$, which we then represent as the Mellin-Barnes integral using the following identity 
\be
\label{MB_pFq}
\ba{l}
\dps
_{p}F_{q}(a_1,...,a_p;b_1,...,b_q|x) = \sum_{n=0}^\infty \frac{(a_1)_n...(a_p)_n}{(b_1)_n,...,(b_q)_n}\frac{x}{n!}
\vspace{2.5mm}  
\\
\dps
\hspace{32mm}
= \frac{1}{2i \pi }\frac{\Gamma(b_1)...\Gamma(b_q)}{\Gamma(a_1)...\Gamma(a_p)}\int_{-i \infty}^{i \infty} dc\; \frac{\Gamma(a_1 + c)...\Gamma(a_p + c)\Gamma(-c)}{\Gamma(b_1 + c)...\Gamma(b_q + c)} (-x)^c\,.
\ea
\ee 
\begin{figure}
\centering
\includegraphics[scale=0.9]{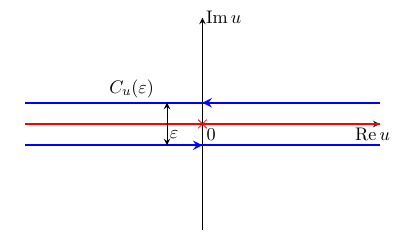}
\caption{The contour $C_u$ on the complex $u$-plane. The red cross and the red line are respectively the singularity and the branch cut of the integrand \eqref{interm_G_tilde}; the  regularization parameter  $\varepsilon \to 0$.}
\label{fig:contour_Cu}
\end{figure}
The resulting expression takes the form 
\be 
\ba{l}
\label{interm_G_tilde}
\dps
V(z_1,z_2,\bx_3)=  \frac{(-)^{h_3-2h_2-2h_3}}{2i\pi^2}\oint_{|u|>u_3} \frac{du}{u}\;\int_{-i \infty}^{i \infty} dc\;\frac{\Gamma(-c)\Gamma(c+1)}{(\dth_{12|c}-\h_3)(\dth_{12|c}+\h_3-1)} 
\vspace{2.5mm}  
\\
\dps
\times
\sum_{\substack{k_1,...,k_4=0\\l_1,...,l_4=0}}^{\infty}\frac{(\dth_{12|c})_{L}}{(2\dth_{12|c})_{L}}\frac{(1-\dth_{12|c})_{K}}{(2-2\dth_{12|c})_{K}}\frac{(h_1)_{k_1+l_1}(h_1)_{k_2+l_2}(h_2)_{k_3+l_3}(h_2)_{k_4+l_4}}{k_1!k_2!k_3!k_4!l_1!l_2!l_3!l_4!}2^{K+L}u^{K}u_3^{L}\left(-\frac{u_3^2}{u^2}\right)^c
\vspace{2.5mm}  
\\
\dps
\times
(2u+u_3+iz_{13})^{-\h_1-k_1-l_1}(u_3+iz_{13})^{-\h_1-k_2-l_2}
(2u+u_3+iz_{23})^{-\h_2-k_3-l_3}(u_3+iz_{23})^{-\h_2-k_4-l_4}\,.
\ea
\ee 
To simplify the integral over $u$ one replaces  the integration contour by the contour $C_u(\varepsilon)$ shown in fig. \bref{fig:contour_Cu}. This change allows one to rewrite the integral over $u$  as a difference of two integrals, and, therefore, as the integral over the discontinuity of the integrand. A similar procedure was performed  when proving  Lemma \bref{lem:PROPTILDE}, see \eqref{integral_interm_pre} and \eqref{integral_interm}. Repeating this procedure, calculating the discontinuity of the factor $\left(-\frac{u_3^2}{u^2}\right)^c$ using \eqref{disc_power}, taking the limit $\varepsilon\to 0$ and showing that at $|u| > u_3$ the Mellin-Barnes integral is zero, one obtains
\be  
\ba{l}
\dps
V(z_1,z_2,\bx_3)=   -\frac{(-)^{\h_1-2\h_2-2\h_3}}{i \pi} \int_{0}^{u_3} \frac{du}{u}\;\int_{-i\infty}^{i\infty} dc\; \frac{1}{(\dth_{12|c}-\h_3)(\dth_{12|c}+\h_3-1)} 
\vspace{2.5mm}  
\\
\dps
\times
\sum_{\substack{k_1,...,k_4=0\\l_1,...,l_4=0}}^{\infty}\frac{(\dth_{12|c})_{L}}{(2\dth_{12|c})_{L}}\frac{(1-\dth_{12|c})_{K}}{(2-2\dth_{12|c})_{K}}\frac{(h_1)_{k_1+l_1}(h_1)_{k_2+l_2}(h_2)_{k_3+l_3}(h_2)_{k_4+l_4}}{k_1!k_2!k_3!k_4!l_1!l_2!l_3!l_4!}2^{K+L}u^{K - 2c}u_3^{L + 2c}
\ea
\ee 
$$
\dps
\times
(2u+u_3+iz_{13})^{-\h_1-k_1-l_1}(u_3+iz_{13})^{-\h_1-k_2-l_2}
(2u+u_3+iz_{23})^{-\h_2-k_3-l_3}(u_3+iz_{23})^{-\h_2-k_4-l_4}\,,
$$
where we also used that the integrand is even function of $u$. Making the change $l_i=n_i-k_i$, $k_1=K_1-k_2-k_3-k_4$ and summing over $k_2,k_3$, $k_4$ by means of the following identity \cite{Bateman:100233}
\be 
\label{2F1_in_1}
\sum_{k=0}\frac{(a)_k(b)_k}{k!(c)_k} = \frac{\Gamma(c)\Gamma(c-a-b)}{\Gamma(c-a)\Gamma(c-b)}\,,
\qquad \text{Re}(c-a-b)>0\,,
\ee 
one obtains 
\be  
\ba{l}
\dps
V(z_1,z_2,\bx_3)=  -\frac{(-)^{\h_1-2\h_2-2\h_3}}{i \pi}\int_{0}^{u_3} \frac{du}{u}\;\int_{-i\infty}^{i\infty} dc\; \frac{1}{(\dth_{12|c}-\h_3)(\dth_{12|c}+\h_3-1)}
\vspace{2.5mm}  
\\
\dps
\times
 \sum_{n_1,...,n_4=0}^{\infty}\frac{(h_1)_{n_1}(h_1)_{n_2}(h_2)_{n_3}(h_2)_{n_4}}{n_1!n_2!n_3!n_4!}\sum_{K_1=0}^{N}\frac{(1-\dth_{12|c})_{N}(\dth_{12|c})_{N-K_1}}{(2-2\dth_{12|c})_{K_1}(2\dth_{12|c})_{N-K_1}}\frac{(N)!}{K_1!(N-K_1)!}
\ea
\ee 
$$
\ba{l}
\dps
\hspace{-37mm}\times
2^N u^{K_1-2c} u_3^{N-K_1+2c} (2u+u_3+iz_{13})^{-\h_1-n_1}(u_3+iz_{13})^{-\h_1-n_2}
\vspace{2.5mm}  
\\
\dps
\hspace{-37mm}\times
(2u+u_3+iz_{23})^{-\h_2-n_3}(u_3+iz_{23})^{-\h_2-n_4}\,,
\ea
$$
where $N = n_1+n_2+n_3+n_4$. Applying the following identity to the sum over $K_1$ \cite{Bateman:100233}
\be 
\label{3F2_identity}
\dps
\sum_{k=0}^\infty\frac{(a)_k(b)_k(c)_k}{k!(a-b+1)_k(a-c+1)_k}z^k = (1-z)^{-a}\sum_{k=0}^\infty\frac{(a-b-c+1)_k(\frac{a}{2})_k(\frac{a}{2}+\half)_k}{k!(a-b+1)_k(a-c+1)_k}\left(-\frac{4z}{(1-z)^2}\right)^k\,,
\ee
where $|z|<1$ and $|\frac{4z}{(1-z)^2}|<1$, one obtains
\be 
\ba{l}
\dps
V(z_1,z_2,\bx_3)=  -\frac{(-)^{\h_1-2\h_2-2\h_3}}{i \pi}\int_{0}^{u_3} \frac{du}{u}\;\int_{-i\infty}^{i\infty} dc\; \frac{1}{(\dth_{12|c}-\h_3)(\dth_{12|c}+\h_3-1)}
\vspace{2.5mm}  
\\
\dps
\times
 \sum_{n_1,...,n_4, K_1=0}^{\infty}\frac{(h_1)_{n_1}(h_1)_{n_2}(h_2)_{n_3}(h_2)_{n_4}}{K_1!n_1!n_2!n_3!n_4!}\frac{(1-\dth_{12|c})_{N}(\dth_{12|c})_{N-K_1}}{(2-2\dth_{12|c})_{K_1}(2\dth_{12|c})_{N-2K_1}}(-)^{K_1}
\vspace{2.5mm}  
\\
\dps
\times
2^N u^{K_1-2c}u_3^{1 -\dth_{12|c}+K_1}(u_3+u)^{2\dth_{12|c}+N-2K_1-1}(2u+u_3+iz_{13})^{-\h_1-n_1}(u_3+iz_{13})^{-\h_1-n_2}
\vspace{2.5mm}  
\\
\dps
\times
(2u+u_3+iz_{23})^{-\h_2-n_3}(u_3+iz_{23})^{-\h_2-n_4}\,.
\ea
\ee 

Now, using \eqref{series_Lauricella} and \eqref{def_lauricella}, one rewrites the sums over $n_i$ as the integral over $v$:
\be 
\ba{l}
\dps
V(z_1,z_2,\bx_3)=   \frac{1}{i \pi}\int_{0}^{u_3} \frac{du}{u}\;\sum_{K=0}^{\infty} \int_{-i\infty}^{i\infty} dc\;\left(\frac{v(1-v)(u_3+u)^2}{u u_3}\right)^{2c}\frac{u^{K}u_3^{1-h_1-h_2+K}(\dth_{12|c} - 1)}{(\dth_{12|c}-\h_3)(\dth_{12|c}+\h_3-1)}
\vspace{2.5mm}  
\\
\dps
\frac{(u_3+u)^{2h_1+2h_2-1-2K}}{K!\Gamma(\dth_{12|c}-K)^2\Gamma(2-2\dth_{12|c}+K)}
\int_0^1dv\;(v(1-v))^{h_1+h_2-K-1}
\vspace{2.5mm}  
\\
\dps
\times
(u^2 - (u+u_3+iz_{13}-2v(u+u_3))^2)^{-\h_1}(u^2 - (u+u_3+iz_{23}-2v(u+u_3))^2)^{-\h_2}\,.
\ea
\ee 
To calculate the Mellin-Barnes integral over $c$ one notes that it should be closed to the left, otherwise the integral is zero as there are no poles to the right of the contour. This leads to the restriction ${v(1-v)(u_3+u)^2} \geq {u u_3}$, which is equivalent to restricting the integration domain in the second integral: $v\in[\frac{u}{u+u_3},\frac{u_3}{u+u_3}]$.  Closing the contour to the left, one calculates the residues of simple poles located at $2c = \h_1-\h_2-\h_3$ and $2c = 1-\h_1-\h_2-\h_3$, makes the change $v=\frac{u_3+u-iz+iz_3}{2(u+u_3)}$ and then sums over $K_1$:
\be 
\label{V_interm}
\ba{l}
\dps
V(z_1,z_2,\bx_3)= \int_{0}^{u_3} \frac{du}{u^2}u^{h_1+h_2}\int_{z_3 - i (u_3-u)}^{z_3 + i (u_3-u)}dz\;(u^2+(z_1-z)^2)^{-\h_1}(u^2+(z_2-z)^2)^{-\h_2}
\vspace{2.5mm}  
\\
\dps
\times
\left[\frac{\Gamma(2h_3-1)}{\Gamma(h_3)^2}\left(\frac{4u u_3}{(u+u_3)^2+(z_3-z)^2}\right)^{1-h_3}\F\left(1-h_3,1-h_3;2-2h_3\Big|\frac{4u u_3}{(u+u_3)^2+(z_3-z)^2}\right)\right.
\vspace{2.5mm}  
\\
\dps
+\frac{\Gamma(1-2h_3)}{\Gamma(1-h_3)^2}\left.\left(\frac{4u u_3}{(u+u_3)^2+(z_3-z)^2}\right)^{h_3}\F\left(h_3,h_3;2h_3\Big|\frac{4u u_3}{(u+u_3)^2+(z_3-z)^2}\right)\right].
\ea
\ee 
The sum of hypergeometric functions in the last two lines can be transformed into a single hypergeometric function using the following identity \cite{Bateman:100233}
\be 
\label{2f1_exp_1}
\ba{l}
\dps
\frac{\Gamma(2a-1)}{\Gamma(a)^2}x^{1-a}\F\left(1-a,1-a;2-2a\Big|x\right)+x^a\frac{\Gamma(1-2a)}{\Gamma(1-a)^2}\F\left(a,a;2a\Big|x\right)
\vspace{2.5mm}  
\\
\dps
=x^a\F\left(a,a;1\Big|1-x\right) = -\left(\frac{x}{2-x}\right)^{1-a}\F\left(1-\frac{a}{2},\half-\frac{a}{2};1\Big|\frac{4(1-x)}{(2-x)^2}\right),
\ea
\ee 
where we also used the quadratic identity in the last line. Using   \eqref{2f1_exp_1} in \eqref{V_interm} yields 
\be 
\label{KGK_tilde_final}
\ba{l}
\dps
V(z_1,z_2,\bx_3)= 
\int_{0}^{u_3} \frac{du}{u^2}u^{h_1+h_2}\int_{z_3 - i (u-u_3)}^{z_3 + i (u-u_3)}dz\;(u^2+(z_1-z)^2)^{-\h_1}(u^2+(z_2-z)^2)^{-\h_2}
\vspace{2.5mm}  
\\
\dps
\times
\left(\frac{2u u_3}{u^2+u_3^2+(z_3-z)^2}\right)^{1-h_3}\F\left(1-\frac{h_3}{2},\half-\frac{h_3}{2};1\Big|1-\left(\frac{2u u_3}{u^2+u_3^2+(z_3-z)^2}\right)^2\right) 
\vspace{2.5mm}  
\\
\dps
= \frac{4^{\h_3}}{-2i}\frac{\Gamma(\h_3)\Gamma(\h_3)}{\Gamma(2\h_3)}\int_{0}^{u_3} \frac{du}{u^2}\int_{z_3 - i (u-u_3)}^{z_3 + i (u-u_3)}dz\; K_{\h_1}(\bx,z_1)K_{\h_2}(\bx,z_2)\widetilde{G}_{h_3}(\bx,\bx_3)\,,
\ea
\ee 
where we used the modified propagator $\widetilde{G}_{h_3}(\bx,\bx_3)$ \eqref{propagators}
\be 
\ba{l}
\dps
\widetilde{G}_{h}(\bx,\bx') = \frac{-2i}{4^{\h}}\frac{\Gamma(2\h)}{\Gamma(\h)\Gamma(\h)}\left(\frac{\Gamma(1-2h)}{\Gamma(1-h)^2} G_h(\bx,\bx') + \frac{\Gamma(2h-1)}{\Gamma(h)^2} G_{1-h}(\bx,\bx')\right) 
\vspace{2.5mm}  
\\
\dps
=\frac{-2i}{4^{\h}}\frac{\Gamma(2\h)}{\Gamma(\h)\Gamma(\h)}\xi(\bx,\bx')^{1-h}\F\left(1-\frac{h}{2},\half-\frac{h}{2};1\Big|1-\xi(\bx,\bx')^2\right).
\ea
\ee
This proves the Lemma \bref{lem:DOUBLE-TRACE} in the case when the points $\bx_1$ and  $\bx_2$ are located on the boundary. In the case when these  points are in the bulk, one multiplies the relation \eqref{KGK_tilde_final} by two smearing functions $\mathbb{K}_{h_1}(\bx_1,w_1)$, $\mathbb{K}_{h_2}(\bx_2,w_2)$ \eqref{smear}, integrates over the boundary points $w_1$, $w_2$ and uses the conversion identity \eqref{KK_as_bb}, which turns the bulk-to-boundary propagators in \eqref{KGK_tilde_final} into the bulk-to-bulk propagators. These steps provide the final proof of Lemma \bref{lem:DOUBLE-TRACE}.  Also note that the restrictions $|z_{13}|>2u_3$ and $|z_{23}|>2u_3$ imposed on $z_1,z_2, z_3$ can be lifted by means of analytic continuation, as both sides of the identity \eqref{KGK_tilde_final} are well-defined in the domain $|z_{13}|<2u_3$ and $|z_{23}|<2u_3$ as long as $z_i\neq z_j, \forall i\neq j$. 

Note that the right-hand side of \eqref{G_tilde_as_vertex} can be rewritten using the conversion identity \eqref{KK_as_bb}, the integral representation of the $3$-point conformal correlation function \eqref{3pt_poch}, and the holographic reconstruction formula \eqref{3ptVertex}:
\be 
\ba{c}
\dps
\pi\int_{0}^{u_3} \frac{du}{u^2} \int_{z_3-i(u-u_3)}^{z_3+i(u-u_3)}\hspace{-2mm}dz\; G_{\h_1}(\bx,\bx_1) G_{\h_2}(\bx,\bx_2) \widetilde{G}_{h_3}(\bx,\bx_3)
\vspace{2.5mm}  
\\
\dps
=\sum_{n=0}^{\infty}\frac{a(\h_3;\h_1,\h_2;n)}{\pref_{\h_1\h_2\dth_{12|n}}}\cV_{ \h_1\h_2 \dth_{12|n}}(\bx_1,\bx_2,\bx_3)\,,
\ea
\ee 
which proves Corollary \bref{coro}.

\subsection{Lemma \bref{lem:PROPID}}
\label{app:lemma4}

Let us  consider the following integral: 
\be 
\label{prop_id_app}
\ba{l}
\dps
\lim_{\varepsilon\to0}\int_{z_i-iu_i}^{z_i+iu_i}dw_i\int_{0}^{\infty} \frac{du}{u^2}\int_{C(\varepsilon)+C_i(\varepsilon)}dz\;\widehat{G}_{h_i}(\bx,\bx_i,w_i)f(\bx)\,,
\ea
\ee 
where the contour $C_i(\varepsilon)$ is shown in fig. \bref{fig:C_i}, the sum $C(\varepsilon)+C_i(\varepsilon)$ is a continuous curve, and $C(\varepsilon)$ is any contour on the complex $z$-plane that avoids  possible  poles of $f(\bx)$ while satisfying  $\re(z)\neq z_i$, $\forall z\in C(\varepsilon)$. 

The proof is straightforward: one splits the integral over $z$ in \eqref{prop_id_app} into two integrals over contours $C(\varepsilon)$ and $C_i(\varepsilon)$, then applies Lemma \bref{lem:PROPSIMPLE} to the integral over $C(\varepsilon)$ and Lemma \bref{lem:PROPTILDE} to the integral over $C_i(\varepsilon)$: 
\be 
\label{prop_id_app_end}
\ba{l}
\dps
\lim_{\varepsilon\to0}\int_{0}^{\infty} \frac{du}{u^2}\int_{C(\varepsilon)}dz\;G_{\h_i}(\bx,\bx_i)f(\bx) = \pi\int_{0}^{u_i} \frac{du}{u^2}\int_{z_i-i(u-u_i)}^{z_i+i(u-u_i)}dz\;\widetilde{G}_{h_i}(\bx,\bx_i)f(\bx)
\vspace{2.5mm}  
\\
\dps
\hspace{32mm}+\lim_{\varepsilon\to0}\int_{z_i-iu_i}^{z_i+iu_i}dw_i\int_{0}^{\infty} \frac{du}{u^2}\int_{C(\varepsilon)+C_i(\varepsilon)}dz\;\widehat{G}_{h_i}(\bx,\bx_i,w_i)f(\bx)\,,
\ea
\ee 
where $f(\bx)$ is subject to the same restriction as in Lemma \bref{lem:PROPTILDE}. This proves Lemma \bref{lem:PROPID}.

Note that a similar relation between the bulk-to-bulk and modified propagators  in Loren\-tzi\-an AdS$_2$  was  explicitly found in the $h_i= 1$ case in  \cite{Kabat:2011rz} (the authors also discussed a generalization to any $h_i$)
\be 
\label{Kabat_rel}
G_h(\bx,\bx') = \pi\, \theta(u'-u)\theta(u'-u-|z'-z|)\,\widetilde{G}_h(\bx,\bx') + i\int_{z'-u'}^{z'+u'}dw\;\widehat{G}_h(\bx,\bx',w)\,,
\ee 
where $\theta(x)$ is the Heaviside step function. This relation is similar to the conversion identity \eqref{KK_as_bb} in a domain where the Heaviside step functions vanish 
\be 
\label{Kabat_rel_null}
G_h(\bx,\bx') = i\int_{z'-u'}^{z'+u'}dw\;\widehat{G}_h(\bx,\bx',w)\,,\quad u > u'-|z'-z|\,,
\ee 
cf. \eqref{KK_as_bb}. By changing $z'\to iz'$, $z\to iz$, $w\to iw$ and requiring $z,z'\in\mathbb{R}$ in \eqref{KK_as_bb} one obtains \eqref{Kabat_rel_null}. 

On the other hand, the superposition identity \eqref{prop_id} can be obtained from \eqref{Kabat_rel}, although with some limitations. To this end, we integrate the identity \eqref{Kabat_rel} with a test function over $\bx\in\RR_+\times\RR$, introduce the $\varepsilon$-prescription in the bulk-to-bulk propagator to avoid crossing its branch cuts and then make the Wick rotation $z'\to iz'$:
\be 
\label{Kabat_rel_int}
\ba{l}
\dps
\int_{0}^{\infty} \frac{du}{u^2}\int_{-i\infty}^{i\infty}dz \;G_h(\bx,\bx')f(\bx) = \pi\, \int_{0}^{u'} \frac{du}{u^2}\int_{z'+i(u'-u)}^{z'-i(u'-u)}dz \;\widetilde{G}_h(\bx,\bx')f(\bx)
\vspace{2.5mm}  
\\
\dps
\hspace{41mm}
+\int_{0}^{\infty} \frac{du}{u^2}\int_{-i\infty}^{i\infty}dz \;\int_{z'-iu'}^{z'+iu'}dw\;\widehat{G}_h(\bx,\bx',w)f(\bx)\,,
\ea
\ee 
where we also changed the integration variables $z\to iz$ and $w\to iw$. The only difference between this  identity and the superposition identity \eqref{prop_id} is the integration contour over $z$. By rotating it around the point $z=0$ by $\frac{\pi}{2}$ one transforms this contour into the line contour $\RR$, which corresponds to choosing $C(\varepsilon) = \RR\symbol{92}U_{\varepsilon}(z')$ in the superposition identity \eqref{prop_id}. Note that this rotation implies that a test  function $f(\bx)$ is holomorphic  in $z\in\CC$, whereas the superposition identity allows $f(\bx)$ to have singularities outside the integration domain.

\subsection{Lemma \bref{lem:propid_n}}
\label{app:coroll2}

Let us  consider the following integral
\be 
\int_{0}^{\infty} \frac{du}{u^2} \int_{\RR}dz\;\prod_{i=1}^n G_{\h_i}(\bx,\bx_i) f(\bx)
= \int_{0}^{\infty} \frac{du}{u^2} \int_{\RR}dz\; G_{\h_1}(\bx,\bx_1) g_1(\bx)\,,
\ee 
where $f(\bx)$ on the left-hand side  is a test function and the product of $n-1$ bulk-to-bulk propagators and the test function $f(\bx)$ on the right-hand side is denoted as $g_1(\bx)$. Now, we apply the superposition identity \eqref{prop_id} to the bulk-to-bulk propagator $G_{\h_1}(\bx,\bx_1)$ with the contour $C(\varepsilon_1)$  taken as $\RR\symbol{92}U_{\varepsilon_1}(z_1)$:
\be 
\label{interm_nprop}
\ba{l}
\dps
\lim_{\varepsilon_1\to0}  \int_{0}^{\infty} \frac{du}{u^2} \int_{\RR\symbol{92}U_{\varepsilon_1}(z_1)}dz\; G_{\h_1}(\bx,\bx_1) g_1(\bx) = \pi\int_{0}^{u_1} \frac{du}{u^2}\int_{z_1-i(u-u_1)}^{z_1+i(u-u_1)}dz\;\widetilde{G}_{h_1}(\bx,\bx_1)g_1(\bx)
\vspace{2.5mm}  
\\
\dps
\hspace{32mm}+\lim_{\varepsilon_1\to0}\int_{z_1-iu_1}^{z_1+iu_1}dw_1\int_{0}^{\infty} \frac{du}{u^2}\int_{\RR\symbol{92}U_{\varepsilon_1}(z_1)+C_1(\varepsilon_1)}dz\;\widehat{G}_{h_1}(\bx,\bx_1,w_1)g_1(\bx)\,.
\ea
\ee 
Since $g_1(\bx)$ contains the product of bulk-to-bulk propagators, one can again apply the superposition identity  to one of the propagators inside $g_1(\bx)$. Rewriting the second term in \eqref{interm_nprop}, one obtains 
\be 
\label{second_term}
\ba{l}
\dps
\lim_{\varepsilon_1\to0}\int_{z_1-iu_1}^{z_1+iu_1} dw_1 \int_{0}^{\infty} \frac{du}{u^2}\int_{\RR\symbol{92}U_{\varepsilon_1}(z_1)+C_1(\varepsilon_1)}dz\; \widehat{G}_{h_1}(\bx,\bx_1,w_1)g_1(\bx) 
\vspace{2.5mm}  
\\
\dps
=  \lim_{\varepsilon_1\to0}\int_{z_1-iu_1}^{z_1+iu_1}dw_1\int_{0}^{\infty} \frac{du}{u^2}\int_{\RR\symbol{92}U_{\varepsilon_1}(z_1)+C_1(\varepsilon_1)}dz\;G_{\h_2}(\bx,\bx_2) g_2(\bx, w_1) \,,
\ea
\ee 
where we denoted the product of $n-2$ bulk-to-bulk propagators and the modified propagator $\widehat{G}_{h_1}$ as $g_2(\bx, w_1)$.

To apply the superposition identity to the bulk-to-bulk propagator $G_{\h_2}(\bx,\bx_2)$ one chooses the contour $C(\varepsilon_2)$ as  $(\RR\symbol{92}U_{\varepsilon_1}(z_1)+C_1(\varepsilon_1))\symbol{92}U_{\varepsilon_2}(z_2)$. Since   $z_1\neq z_2$, $\exists \varepsilon$ such that for  $\forall\varepsilon_1< \varepsilon$ and $\forall \varepsilon_2 < \varepsilon$: $U_{\varepsilon_2}(z_2)\cap C_1(\varepsilon_1)= \varnothing$. As $\varepsilon_{1,2} \to 0$, we assume that $\varepsilon_{1,2} < \varepsilon$, and, therefore, the contour $C(\varepsilon_2)$ is given by the contour $C(\varepsilon_1,\varepsilon_2) \equiv  \RR\symbol{92}(U_{\varepsilon_1}(z_1)\cup U_{\varepsilon_2}(z_2))+C_1(\varepsilon_1)$. The result of applying  the superposition identity to \eqref{second_term} is given by  
\be 
\label{prod_exp_interm}
\ba{l}
\dps
\lim_{\varepsilon_2\to 0}\int_{0}^{\infty} \frac{du}{u^2} \int_{C(\varepsilon_1,\varepsilon_2)}dz \; G_{\h_2}(\bx,\bx_2) g_2(\bx, w_1)
=  \pi\int_{0}^{u_2} \frac{du}{u^2}\int_{z_2-i(u-u_2)}^{z_2+i(u-u_2)}dz\;\widetilde{G}_{h_2}(\bx,\bx_2) g_2(\bx, w_1)
\vspace{2.5mm}  
\\
\dps
+\lim_{\varepsilon_2\to 0}\int_{z_2-iu_2}^{z_2+iu_2}dw_2 \int_{0}^{\infty}\frac{du}{u^2} \int_{C(\varepsilon_1,\varepsilon_2)+C_2(\varepsilon_2)}dz\; \widehat{G}_{\h_2}(\bx,\bx_2,w_2) g_2(\bx, w_1)\,.
\ea
\ee 
Repeating this procedure  $n-2$ times and then applying the conversion identity \eqref{KK_as_bb} to the bulk-to-bulk propagators in the terms with $\widetilde{G}$ one obtains the multipoint superposition identity \eqref{propid_n}.

\bibliographystyle{JHEP}
\providecommand{\href}[2]{#2}\begingroup\raggedright\endgroup

\end{document}